\documentstyle[epsf,twocolumn,prd,aps,floats]{revtex}
\begin{document}

\draft
\date{September 20, 2000}
%
%
\newcommand{\nc}{\newcommand}
\nc{\bea}{\begin{eqnarray}}
\nc{\eea}{\end{eqnarray}}
\nc{\beq}{\begin{equation}}
\nc{\eeq}{\end{equation}}
\nc{\bi}{\begin{itemize}}
\nc{\ei}{\end{itemize}}
\nc{\la}[1]{\label{#1}}

\nc{\half}{\frac{1}{2}}

\nc{\p} {{\rm p}}
\nc{\n} {{\rm n}}
\nc{\A} {{\rm A}}
\nc{\D}{{\rm D}}
\nc{\EH}{{}^3{\rm H}}
\nc{\EHe}{{}^3{\rm He}}
\nc{\UHe}{{}^4{\rm He}}
\nc{\GLi}{{}^6{\rm Li}}
\nc{\ZLi}{{}^7{\rm Li}}
\nc{\ZBe}{{}^7{\rm Be}}
\nc{\DH}{{\rm D}/{\rm H}}
\nc{\EHeH}{^3{\rm He}/{\rm H}}
\nc{\ZLiH}{{}^7{\rm Li}/{\rm H}}

\nc{\nbar}  {\bar {\rm n}}
\nc{\pbar}  {\bar {\rm p}}
\nc{\Abar}  {\bar {\rm A}}
\nc{\dbar}  {\overline {\rm D}}
\nc{\tbar}  {{}^3\overline {\rm H}}
\nc{\EHebar}{{}^3\overline {\rm He}}
\nc{\UHebar}{{}^4\overline {\rm He}}

\nc{\pHe} {\pbar\UHe}
\nc{\pHebar} {\p\UHebar}
\nc{\nHe} {\nbar\UHe}
\nc{\nHebar} {\n\UHebar}

\nc{\et}{\eta_{10}}
\nc{\GeV}{\mbox{ GeV}}
\nc{\MeV}{\mbox{ MeV}}
\nc{\keV}{\mbox{ keV}}
\nc{\eV}{\mbox{ eV}}
\nc{\etal}{{\it et al.}}
\nc{\x}[1]{}
\nc{\im}{\mbox{Im}}
\nc{\imasc}{\mbox{Im}(-a_{\rm sc})}

\nc{\nenet}{n_e^\ast}
\nc{\nbref}{{\langle n_B \rangle}}
\nc{\netot}{n_e}
\nc{\nbtot}{n_B}
\nc{\nbbar}{n_{\bar{b}}}
\nc{\msq}{{\rm m}^2}
\nc{\me}{m_{\rm e}}
\nc{\ve}[1]{{\bf #1}}

%
%

\nc{\AJ}[3]{{Astron.~J.\ }{{\bf #1}{, #2}{ (#3)}}}
\nc{\anap}[3]{{Astron.\ Astrophys.\ }{{\bf #1}{, #2}{ (#3)}}}
\nc{\ApJ}[3]{{Astrophys.~J.\ }{{\bf #1}{, #2}{ (#3)}}}
\nc{\apjl}[3]{{Astrophys.~J.\ Lett.\ }{{\bf #1}{, #2}{ (#3)}}}
\nc{\app}[3]{{Astropart.\ Phys.\ }{{\bf #1}{, #2}{ (#3)}}}
\nc{\araa}[3]{{Ann.\ Rev.\ Astron.\ Astrophys.\ }{{\bf #1}{, #2}{ (#3)}}}
\nc{\arns}[3]{{Ann.\ Rev.\ Nucl.\ Sci.\ }{{\bf #1}{, #2}{ (#3)}}}
\nc{\arnps}[3]{{Ann.\ Rev.\ Nucl.\ and Part.\ Sci.\ }{{\bf #1}{, #2}{ (#3)}}}
\nc{\epj}[3]{{Eur.\ Phys.\ J.\ }{{\bf #1}{, #2}{ (#3)}}}
\nc{\MNRAS}[3]{{Mon.\ Not.\ R.\ Astron.\ Soc.\ }{{\bf #1}{, #2}{ (#3)}}}
\nc{\mpl}[3]{{Mod.\ Phys.\ Lett.\ }{{\bf #1}{, #2}{ (#3)}}}
\nc{\Nat}[3]{{Nature }{{\bf #1}{, #2}{ (#3)}}}
\nc{\ncim}[3]{{Nuov.\ Cim.\ }{{\bf #1}{, #2}{ (#3)}}}
\nc{\nast}[3]{{New Astronomy }{{\bf #1}{, #2}{ (#3)}}}
\nc{\np}[3]{{Nucl.\ Phys.\ }{{\bf #1}{, #2}{ (#3)}}}
\nc{\pr}[3]{{Phys.\ Rev.\ }{{\bf #1}{, #2}{ (#3)}}}
\nc{\PRC}[3]{{Phys.\ Rev.\ C\ }{{\bf #1}{, #2}{ (#3)}}}
\nc{\PRD}[3]{{Phys.\ Rev.\ D\ }{{\bf #1}{, #2}{ (#3)}}}
\nc{\PRL}[3]{{Phys.\ Rev.\ Lett.\ }{{\bf #1}{, #2}{ (#3)}}}
\nc{\PL}[3]{{Phys.\ Lett.\ }{{\bf #1}{, #2}{ (#3)}}}
\nc{\prep}[3]{{Phys.\ Rep.\ }{{\bf #1}{, #2}{ (#3)}}}
\nc{\RMP}[3]{{Rev.\ Mod.\ Phys.\ }{{\bf #1}{, #2}{ (#3)}}}
\nc{\rpp}[3]{{Rep.\ Prog.\ Phys.\ }{{\bf #1}{, #2}{ (#3)}}}
\nc{\zphysA}[3]{{Z.\ Phys.\ A }{{\bf #1}{, #2}{ (#3)}}}
\nc{\ibid}[3]{{\it ibid.\ }{{\bf #1}{, #2}{ (#3)}}}

\nc{\pla}[3]{{Plasma Phys.\ }{{\bf #1}{, #2}{ (#3)}}}
\nc{\ndt}[3]{{Nuclear Data Tables\ }{{\bf #1}{, #2}{ (#3)}}}

\wideabs{
\title{Antimatter Regions in the Early Universe and Big Bang Nucleosynthesis}

\author{Hannu Kurki-Suonio\cite{mailh}}
\address{Helsinki Institute of Physics,
         P.O.Box 9, FIN-00014 University of Helsinki, Finland}

\author{Elina Sihvola\cite{maile}}
\address{Department of Physics, University of Helsinki,
         P.O.Box 9, FIN-00014 University of Helsinki, Finland}


\maketitle

\begin{abstract}
We have studied big bang nucleosynthesis in the presence of
regions of antimatter.
Depending on the distance scale of the antimatter region, and thus
the epoch of their annihilation, the amount of antimatter in the
early universe is constrained by the observed abundances.  Small
regions, which annihilate after weak freezeout but before
nucleosynthesis, lead to a reduction in the $\UHe$ yield, because
of neutron annihilation. Large regions, which annihilate after
nucleosynthesis, lead to an increased $\EHe$ yield.  Deuterium
production is also affected but not as much.  The three most
important production mechanisms of $\EHe$ are (1) photodisintegration
of $\UHe$ by the annihilation radiation, (2) $\pHe$ annihilation,
and (3) $\nHe$ annihilation by ``secondary" antineutrons produced in $\UHebar$
annihilation.  Although $\pHe$ annihilation produces more $\EHe$
than the secondary $\nHe$ annihilation, the products of the latter
survive later annihilation much better, since they are distributed
further away from the annihilation zone.
Our results are in qualitative agreement with similar work by 
Rehm and Jedamzik, but we get a larger $\EHe$ yield.

\end{abstract}

 }

\vspace*{-10.3cm}
\noindent \hspace*{15cm} \mbox{HIP-2000-31/TH}
\vspace*{8.3cm}

%
%

\section{Introduction}

\subsection{Antimatter in the Universe}

The local universe is baryon asymmetric.  It contains matter, not
antimatter. In standard homogeneous big bang cosmology the
universe was filled with a uniform mixture of antimatter and
matter very early on, with a slight
excess of matter over antimatter.  This excess of matter was left
over when matter and antimatter annihilated during the first
millisecond.

This baryon asymmetry is characterized by the baryon-to-photon ratio,
\begin{equation}
  \eta
  \equiv \frac{n_b-n_{\bar{b}}}
  {n_\gamma}
  \equiv \frac{n_B}{n_\gamma},
\end{equation}
where $n_b$ is the number density of baryons and $n_{\bar{b}}$ the
number density of antibaryons.

 There are many proposed mechanisms for baryogenesis to
explain the origin of this asymmetry.  The simplest versions of
baryogenesis produce a homogeneous asymmetry, but there are many
possibilities for inhomogeneous baryogenesis, which could produce
a baryon excess in some regions and an antibaryon excess in other
regions. This leads to a structure of matter and antimatter regions 
after local annihilation during the first millisecond.

In scenarios connected with inflation, there is no a priori
constraint on the distance scale of these matter and antimatter
regions. If the distance scale is small, the antimatter regions
would have annihilated in the early universe, and the presence of
matter today requires asymmetric baryogenesis, producing more
baryons than antibaryons.  If the distance
scale is large, antimatter regions would have survived till
present.

In the latter case, an overall baryon symmetry remains a
possibility.  The universe could contain equal amounts of matter and
antimatter, spatially separated into matter and antimatter domains.
In this case, the absence of observed annihilation radiation from the
domain boundaries indicates that the typical size of these domains
would have to be very large.

Considering only the conditions in the present universe, the lower
limit to the domain size corresponds to the scale of cluster of
galaxies, of the order of 20 Mpc\cite{Steigman}.  Because of the
low density of intergalactic space between clusters the
annihilation radiation between a cluster and an ``anticluster''
could have escaped detection.

However, the isotropy of the cosmic microwave background (CMB) rules out
large voids between matter and antimatter regions during an earlier time.
Thus annihilation would have been more intense before structure
formation.  The relic $\gamma$ rays would contribute to the cosmic
diffuse gamma (CDG) spectrum.  The observed CDG spectrum gives a much
larger lower limit to the domain size, of the order of $10^3$ Mpc,
comparable to the size of the visible universe\cite{CRG98}.  A
boundary of an even larger domain intersecting the last scattering
surface could leave an imprint on the CMB\cite{Kinney97}, but these
are unlikely to be observable with planned CMB probes\cite{CR98}.

If we drop the assumption of baryon symmetry, allowing for a
lesser amount of antimatter than matter, then instead of a lower
limit to the domain size, observations just place upper limits to
the antimatter-matter ratio $R$ at different distance scales.
Indeed, it may be possible
to have a small fraction $R < 10^{-6}$ of antimatter stars in
our galaxy\cite{Khlopov}.

 No antinuclei (with $|Z| > 1$) have ever
been observed in cosmic rays. The Alpha Magnetic Spectrometer
(AMS)\cite{AMS} to be placed on the International Space
Station will look for antinuclei in cosmic rays, and if none are
found, will place a tight upper limit on the antimatter
fraction of cosmic ray  sources.  The AMS precursor flight on the
Space Shuttle observed $2.86\times10^6$ helium nuclei but no
antihelium\cite{AMS1}, giving an upper limit $\overline{\rm
He}/{\rm He} < 1.1\times10^{-6}$ on the antihelium-helium flux
ratio in cosmic rays.

\subsection{Inhomogeneous baryogenesis}

Early work on antimatter regions in the universe (see the review
by Steigman\cite{Steigman}) considered them as an initial
condition for the universe\cite{Harrison68}, or tried to form them
by separating matter from antimatter at a later
stage\cite{Omnes}.  Later work is related to
scenarios for inhomogeneous baryogenesis.

There are many proposed mechanisms for baryogenesis, 
including grand unified theory (GUT)
baryogenesis, electroweak baryogenesis, and Affleck-Dine
baryogenesis.  The simplest versions produce a homogeneous
baryoasymmetry, but simple modifications lead to an inhomogeneous
baryogenesis which produces matter and antimatter
regions\cite{Stecker1,Stecker2,CD92,ewbg,GS98}.
See, e.g., the reviews by Dolgov\cite{Dolgovrev}.

Inhomogeneous baryogenesis without inflation leads to a
matter-antimatter domain structure with a very small distance
scale.  Models connected to inflation can lead to arbitrarily
large distance scales.
Some scenarios for GUT baryogenesis lead to an unacceptable large
domain wall energy between the matter and antimatter domains, but
other scenarios avoid this problem\cite{Stecker2}.

Most studies of inhomogeneous baryogenesis have been for a
globally baryon symmetric universe.  As it has been recently
shown\cite{CRG98} that the distance scale in this case would have to
be at least comparable to the present horizon, the attention has
shifted to models where the observable universe is baryon
asymmetric, but could contain a smaller amount of
antimatter\cite{DS93,KRS00}.

\subsection{Antimatter regions in the early universe}

On scales smaller than about 1 kpc\cite{Khlopov00}, antimatter
regions would have annihilated by now but could have left an
observable signature in the CDG spectrum, in the CMB spectrum, 
or in the yields of light elements from big bang nucleosynthesis 
(BBN).

The smaller the size of the antimatter regions, the earlier they
annihilate.  Domains smaller than 100 m at $T=$ 1 MeV, 
corresponding to a comoving (present) scale of $6\times 10^8$ km 
or 0.02 mpc, would annihilate well before nucleosynthesis 
and would leave no observable remnant.

The energy released in antimatter annihilation thermalizes with
the ambient plasma and the background radiation, if the energy
release occurs at $T > 1$ keV.  If the annihilation occurs later,
Compton scattering between electrons heated by the annihilation
and the background photons transfers energy to the microwave
background, but is not able to thermalize this energy (because
Compton scattering conserves photon number).  The lack of observed
distortion in the CMB spectrum constrains the energy release
occurring after $T = 1$ keV to below $6\times10^{-5}$ of the CMB
energy\cite{CMBspectrum}. This leads to progressively
stronger constraints on the amount of antimatter annihilating at
later times, as the ratio of matter and CMB energy density is
getting smaller. Above $T \sim 0.2$ eV the baryonic matter energy
density is smaller than the CMB energy density, so the limits on
the antimatter fraction annihilating then are weaker than
$6\times10^{-5}$.

For scales larger than about 100 pc (or $7\times10^{11}$ m at 1 keV)
the tightest constraints on the amount of antimatter come from the
CMB spectral distortion, and from the CDG spectrum for even larger
scales\cite{CRG98}.

We consider here intermediate distance scales, where most of the
annihilation occurs shortly before or during nucleosynthesis, or
after nucleosynthesis but before recombination, at temperatures
between 1 MeV and 1 eV. The strongest constraints on the amount of
antimatter at these distance scales will come from big bang
nucleosynthesis affected by the annihilation process.

\subsection{BBN with antimatter}

Much of the early work on BBN with
antimatter\cite{Steigman,earlyABBN,Cheche,InjABBN}
was either in the context of a baryon symmetric
universe\cite{earlyABBN} or for a homogeneous injection of
antimatter through some decay process\cite{InjABBN}.

Rehm and Jedamzik\cite{RJ98} studied small antimatter regions,
which annihilate before nucleosynthesis, at temperatures $T >
80\keV$. Because of faster diffusion of neutrons and antineutrons
(as compared to protons and antiprotons), annihilation reduces the
net neutron number, leading to underproduction of
$\UHe$\cite{Steigman}. This sets a limit $R < 10^{-2}$ for the
amount of antimatter in regions of size $r_A \sim 1$ cm at $T=$
100 GeV (2 km at $T = 1\MeV$ or $3\times10^6$ m at $T = 1\keV$).
Our results for these small scales agree with\cite{RJ98}.

We consider also larger antimatter regions, which annihilate
mainly during or after nucleosynthesis. We have done detailed
inhomogeneous nucleosynthesis calculations, where diffusion,
annihilation, and nucleosynthesis all happen simultaneously.

The case where annihilation occurs after nucleosynthesis was
considered in\cite{Cheche}. Because annihilation of
antiprotons on helium would produce D and $\EHe$ it was estimated
that the observed abundances of these isotopes place a comparable
upper limit to the amount of antimatter annihilated after
nucleosynthesis.  As we explain below, the situation is rather
more complicated.

We reported our first results in\cite{KS00}, where we had not
included some features whose effect we estimated to be small.
These included (1) antinucleosynthesis in the antimatter region, (2)
photodisintegration of other isotopes than $\UHe$, and (3) the
dependence of the electromagnetic cascade spectrum on the initial
photon spectrum from annihilation.  We have now made the following
changes to our computer code to take these effects into account.

(1) We have added all antinuclei up to $\bar{A}=4$ and their
antinucleosynthesis.  Annihilation of these antinuclei produce
energetic antimatter fragments which may penetrate deep into the
matter region and annihilate there.  Thus annihilation reactions
occur also far away from the matter-antimatter boundary.

(2) We have added photodisintegration of the lighter nuclei, D,
$\EH$, and $\EHe$.

(3) We treat photodisintegration in more detail, especially at lower
temperatures, where use of the standard cascade spectrum is no
longer appropriate.

Below, all distance scales given in meters will refer to comoving
distance at $T = 1$ keV. One meter at $T=1$ keV corresponds to
$4.24\times10^6$ m or $1.37\times10^{-10}$ pc today. Rehm and
Jedamzik\cite{RJ98} give their distance scales at $T=100$ GeV. Our
distances are thus larger by a factor $3.0\times10^8$. We use
$\hbar=c=k_B=1$ units.

The physics of the annihilation of antimatter regions in the early
universe is discussed in Sec.~II.  We describe our numerical
implementation in Sec.~III and give the results in Sec.~IV.  We
summarize our conclusions in Sec.~V.


\section{Annihilation of Antimatter Domains}

\subsection{Mixing of matter and antimatter}

Consider the evolution of an antimatter region, with radius $r_A$,
surrounded by a larger region of matter.  We are interested in the
period in the early universe when the temperature was between 1
MeV and 1 eV (age of the universe between 1 s and 30000 years).
The universe is radiation dominated during this period. At first
matter and antimatter are in the form of nucleons and
antinucleons, after nucleosynthesis in the form of ions and
anti-ions.  Matter and antimatter are mixed by diffusion at the
boundary and annihilated. Thus there will be a narrow annihilation
zone separating the matter and antimatter regions.

Before nucleosynthesis the mixing of matter and antimatter occurs
mainly through neutron/antineutron diffusion, since neutrons
diffuse much faster than protons.
If the radius of the antimatter region is less than $r\approx10^7$
m, all antimatter annihilates before nucleosynthesis. In
nucleosynthesis the remaining free neutrons go into $\UHe$ nuclei.
The mixing of matter and antimatter practically stops until the
density has decreased enough for ion diffusion  to become
effective at $T\approx 3$ keV.

Thus there are two stages of annihilation, the first one before
nucleosynthesis, at $T\gtrsim70$ keV, the second well after
nucleosynthesis, at $T\lesssim3$ keV.  The physics during the two
regimes is quite different.  The first regime was discussed
in\cite{RJ98}.  We concentrate on the second regime in the
following discussion.

Hydrodynamic expansion becomes important at $T\approx30$ keV. At
that time the annihilation of thermal electron-positron pairs
becomes practically complete and the photon mean free path
increases rapidly. When the mean free path becomes larger than the
distance scale of the baryon inhomogeneity, the baryons stop
feeling the pressure of the photons, which had balanced the
pressure of baryons and electrons. The pressure gradient then
drives the fluid into motion towards the annihilation
zone\cite{Alcock90,JF94a}. This flow is resisted by Thomson drag.
The fluid reaches a terminal velocity \cite{JF94a}
\beq
      v = \frac{3}{4\sigma_T\varepsilon_\gamma
          |\nenet|} \frac{dP}{dr} .
\eeq
Here $\varepsilon_\gamma$ is the energy density of photons,
$\sigma_T=0.665\times10^{-28}\msq$ is the Thomson cross section,
$P$ is the pressure of baryons and electrons, and
$\nenet=n_{e^-}-n_{e^+}$ is the net electron density. With
$P\approx(\nbtot+\netot)T$  and $|\nenet|\approx\nbtot$, we get a
diffusive equation
\beq
     \frac{\partial\nbtot}{\partial t} =
     \nabla\cdot \left(
     \frac{3T}{2\sigma_T\varepsilon_\gamma} \nabla \nbtot
     \right) \label{hydroDE}
\eeq
for the baryon density, with an effective baryon diffusion
constant due to hydrodynamic expansion
\beq
   D_{\rm hyd} = \frac{3T}{2\sigma_T\varepsilon_\gamma} \propto T^{-3} .
\eeq

The hydrodynamic expansion alone does not cause mixing, but it
significantly speeds up annihilation by bringing
material towards the annihilation zone. The annihilation zone is
surrounded by a depletion zone\cite{CRG98}, where the density of
(anti)matter has decreased due to matter flow into the
annihilation zone.  The resulting pressure gradient maintains this
flow.

Antinucleosynthesis in the antimatter region produces antinuclei.
The yields of these anti-isotopes are not interesting in
themselves, since they are eventually annihilated. The
annihilation of these antinuclei with nucleons produces energetic
antinucleons and lighter antinuclei, which may penetrate deep into
the matter region before annihilating.  Thus, in addition to
``primary" annihilation in the annihilation zone, there is also
``secondary" annihilation outside this zone.

\subsection{Annihilation reactions}

The primary annihilation reactions occur at low energies where
reaction data is scarce or non-existent.  Theoretically, the
annihilation cross section is known to behave as $1/v$, when one
or both of the annihilating particles are neutral, and as $1/v^2$
when both are charged.

More precisely, the theoretical $\nbar \A$ cross section
is\cite{Shapiro58,CPZ97}
\beq
   \sigma \approx 4\pi \biggl(\frac{\im(-a_s)}{q} -
   2\im^2(-a_s)\biggr),
\eeq where $a_s$ is the scattering length, $q = \mu v$, $\mu$ is
the reduced mass, and $v$ is the relative velocity of the
annihilating particles.

The analogous expression for S-wave $\pbar \A$ annihilation
is\cite{CP93,CPZ97}
\bea
   \sigma & = & \frac{8\pi^2}{1-\exp(-2\pi\eta)}
   \frac{1}{q^2}
   \frac{\im(-a_{\rm sc}/B)}{|1 + iqw(\eta)a_{\rm sc}|^2} \nonumber \\
   & \approx & C(v)\frac{4\pi}{q}\frac{\im(-a_{\rm sc})}
   {|1+i2\pi a_{\rm sc}/B|^2},
\eea
where $\eta = -1/qB$ is the dimensionless Coulomb parameter,
$B = 1/Z\mu\alpha$ is the Bohr radius of the antiparticle-particle
system, $a_{\rm sc}$ is the Coulomb-corrected scattering length and
\beq
   C(v) \equiv \frac{2\pi Z\alpha/v}{1-\exp(-2\pi Z\alpha/v)}.
\eeq

There is laboratory data only for some of the relevant scattering
lengths, and the uncertainties are large\cite{CP96,Heisenberg}.  From atomic
data\cite{Batty89,CP92,CPZ97}, the $\pbar\p$ system has $\imasc =
0.71 \pm0.05$ fm.
Recent experimental data by the OBELIX group\cite{Protasov99}
gives $\imasc = 0.62\pm0.02\pm0.04$ fm for $\pbar\D$ and
$\imasc = 0.36\pm0.03^{+0.19}_{-0.11}$ for $\pbar\UHe$.

Primary annihilation is not sensitive to the annihilation cross
sections, since annihilation is complete in the annihilation zone
anyway. In secondary annihilation the $A$-dependence of the
annihilation cross section is important, since it determines
whether antinucleons annihilate with protons, which leads to no
nuclear yields, or with $\UHe$, producing $\D$ and $\EHe$.

The yields of the annihilation reactions are important.
Fortunately there is data on the most important reaction,
antiprotons on helium\cite{Balestra88}, and also on some other
reactions with
antiprotons\cite{annyields,Sudov93}.

The annihilation reaction between an antinucleon and a nucleus can be
thought of as an annihilation of one of the nucleons in the nucleus.
According to experimental data, an antiproton is twice as likely to
annihilate on a proton than on a neutron in the
nucleus\cite{Balestra87,Egidy}.

The annihilation of a nucleon and an antinucleon produces a number of
pions, on average 5--6 with a third of them neutral\cite{Steigman,Egidy}.
The charged pions decay into muons and neutrinos, the muons into
electrons and neutrinos.  The neutral pions decay into two photons.
About half of the annihilation energy, 1880 MeV, is carried away by the
neutrinos, one third by the photons, and one sixth by electrons and
positrons.

When an antinucleon annihilates on a nucleus, some of the produced
pions may knock out some of the other nucleons, or in the case of
larger nuclei, small fragments (p, $\D$, $\EH$, $\EHe$, $\UHe$). Some of
the annihilation energy will go into the kinetic energy of these
particles and the recoil energy of the residual nucleus. Typical
energies are of order $\sim 10$ MeV.

According to Balestra et al.\cite{Balestra88} the average yields of
low-energy $\pbar\UHe$ annihilation are $0.210\pm$0.009 $\EHe$,
$0.437\pm$0.032 $\EH$, 0.07--0.19 D. This leaves about 0.7--0.9
nucleons.

Experimental data on the energy spectra of these emitted nucleons
and fragments can be approximated by $Ce^{-E/E_0}$, where the
average energy $E_0$ decreases with the mass of the emitted
particle.  However the corresponding momentum is close to 350
MeV/$c$ independent of mass\cite{Egidy}. The momenta of the
residual nuclei are smaller. Balestra et al.\cite{Balestra88}
report a measurement on the momentum distribution of $\EHe$ from
$\pHe$ annihilation, with the mean energy corresponding to a momentum
of 198 MeV/$c$. Because of their large momenta these reaction
products get spread over a large area, many of them escaping the
annihilation zone, at least for a while.

\subsection{Thermalization of annihilation products}

The annihilation products lose their kinetic energy through
collisions in the ambient plasma. Ions lose energy by Coulomb
scattering on electrons and ions and by Thomson scattering on
photons. If the velocity of the ion is
greater than thermal electron velocities, the energy loss is mainly due to
electrons. At low energies scattering on ions becomes
important.

For an ion with $E\gg T$, we find that the energy loss per unit
distance due to Coulomb collisions is
\bea
     \frac{dE}{dr} =
    && 4\pi n(Zz\alpha)^2\Lambda
        \left( 1+ \frac mM \right) \frac Mm\frac1E \times  \label{Coulomb} \\
      && \Bigg[ \sqrt \frac{mE}{\pi MT}
                \exp \left(-\frac{mE}{MT}\right)
               -\half {\rm erf} \left( {\sqrt\frac{mE}{MT}} \right)
         \Bigg] .    \nonumber
\eea
Here $M$, $Z$, and $E$ are the mass, charge and energy of the incoming ion, $T$
is the temperature of the plasma, $m$, $z$, and $n$  are the mass, charge,
and number density of the plasma particles,
and $\Lambda\sim 15$ is the Coulomb logarithm.
We assumed here that both the incoming ion and the plasma particles are
non-relativistic.

When the ion velocity is large compared to the thermal velocities
of plasma particles, Eq. (\ref{Coulomb}) simplifies into
\cite{Jackson} \beq
     \frac{dE}{dr}  \approx
    -2\pi n(Zz\alpha)^2\Lambda
      \left( 1+ \frac mM \right) \frac Mm \frac 1E,
        \label{Jackson}
\eeq
and in the opposite case, [$mE/(MT)\ll1$], into\cite{Stix}
\beq
     \frac{dE}{dr}  \approx
    -\frac{8\sqrt\pi}{3} n(Zz\alpha)^2\Lambda
      \left( 1+ \frac mM \right) \frac 1T \sqrt{\frac{mE}{MT}}.
        \label{Stix}
\eeq

The energy loss in a plasma consisting of electrons and nuclei is
thus
\bea
     \frac{dE}{dr} =
    && 4\pi n_e(Z\alpha)^2\Lambda_e
       \frac M\me\frac1E \times                    \label{sumions} \\
    && \Bigg[ \sqrt \frac{\me E}{\pi MT}
             \exp \left(-\frac{\me E}{MT}\right)
            -\half {\rm erf}\left(\sqrt\frac{\me E}{MT}\right) \nonumber \\
    &&      -\half \sum_i\left( 1+ \frac{A_i}{A} \right)
            \frac{Z_i^2}{A_i} \frac{\Lambda_i}{\Lambda_e}
            \frac{\me}{m_p} \frac{n_i}{n_e}
                  \Bigg]    \nonumber
\eea
where we assumed $\me/M\ll1$, and used approximation (\ref{Jackson}) for
scattering on nuclei. The index $i$ labels different species of nuclei in
the plasma. 

\begin{figure}[tbh]
\epsfysize=6.8cm
\epsffile{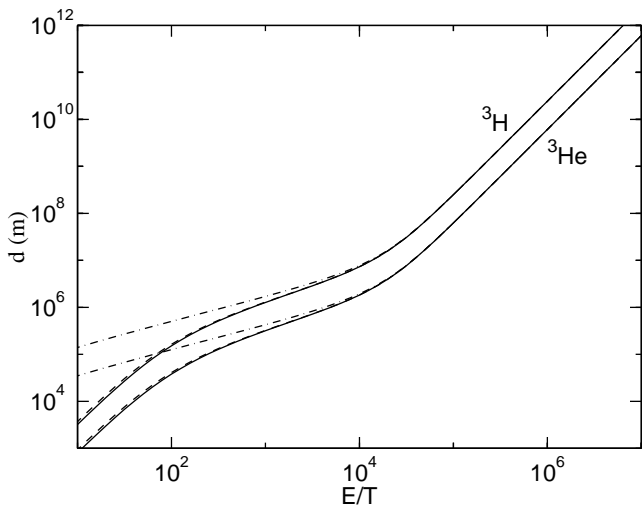}
\caption[a]{\protect
The penetration distance of a $\EHe$ and a $\EH$ ion in
matter with constant baryon density, $\eta=6\times10^{-10}$.
The distance $d$ is given in
comoving units at $T=1$ keV. The {\em  solid line} is for a
case where all baryons are in form of protons,
and the {\em dashed line} for a $\UHe$ mass fraction of
0.25. The {\em dot-dashed line} shows the effect of ignoring
scattering on nuclei.
The penetration distance for an ion with charge $Z$
and mass number $A$ is obtained approximately by scaling the
$\EH$ curve by a factor $A/(3Z^2)$ vertically and by $A/3$ horizontally.
For scattering on electrons (dot-dashed line) this scaling rule is
exact. }
\label{fig:penetration}
\end{figure}

We plot the penetration distance $d$ of $\EHe$ and $\EH$ ions in a
homogeneous plasma in Fig.\ref{fig:penetration}, as a function of
$E/T$, in the absence of thermal electron-positron pairs. At large
energies $d\propto E^2$, which corresponds to approximation
(\ref{Jackson}).
We show also the effect of
ignoring scattering on nuclei.

The drag force exerted on the ion by photons is
\beq
  \frac{dE}{dr} = -\frac43 \left( \frac{\me}{M} \right)^2
                  Z^4   \sigma_T \varepsilon_\gamma v.  \label{ionphoton}
\eeq
The effect of this is negligible compared to Coulomb scattering.

%

Neutrons lose energy through scattering on ions and electrons.
Scattering on electrons is not important for $T<30$ keV.
The neutron loses a substantial part of its energy in each
collision with an ion. The penetration distance is of order of the
mean free path $\lambda=1/(\sigma n)$.   Assuming $\eta = 6\times10^{-10}$,
we find for neutron-proton scattering
$\lambda \approx 4.7\times10^9{\rm~m~}(T/\keV)^{-2}$
for a neutron with a typical 70 MeV energy.
At $T<0.36\keV$ the mean free time of a 70 MeV neutron
becomes larger than its lifetime. The neutron is then likely to
decay into a proton before thermalizing.

\subsection{Photodisintegration}

The high-energy photons and electrons from pion decay initiate
electromagnetic cascades\cite{photo,DEHS,Svensson90,Ellis92,PSB95,KM95,Sigl95}.
 The dominant
thermalization mechanisms for energetic photons and electrons are
photon-photon pair production and inverse Compton scattering
\beq
   \gamma + \gamma_{bb} \rightarrow e^+ + e^-,\quad
   e + \gamma_{bb} \rightarrow e' + \gamma',
\eeq
with the background photons $\gamma_{bb}$.
The cascade proceeds rapidly until the photon energies $E_\gamma$ are below the
threshold  for pair production,
\beq
   E_\gamma \epsilon_\gamma = \me^2,
\eeq
where $\epsilon_\gamma$ is the energy of the background photon.

Because of the large number of background photons, a significant
number of them have energies $\gg T$, and the photon-photon pair
production is the dominant energy loss mechanism for cascade
photons down to\cite{Ellis92}
\beq
   E_{\rm max} = \frac{\me^2}{22T}.
\eeq
Below this threshold energy, the dominant scattering process
is photon-photon scattering down to\cite{Svensson90}
\beq
   E_{\rm c} = \frac{\me^2}{80T}.
\eeq
Below this energy the dominating energy loss processes for
photons are pair production on nuclei and Compton scattering on
electrons. Inverse Compton scattering is still the dominant energy
loss mechanism for electrons.

When energy is released in the form of photons and electrons with energies well
above $E_{\rm max}$, the energy is rapidly converted into a cascade
photon spectrum, which depends only
on the total energy $E_0$ injected, and is well approximated
by\cite{Ellis92,KM95}
\beq
     \frac{dn_\gamma}{dE} =
       \left\{
           \begin{array}{ll}
              A(E/E_c)^{-1.5}, &  E < E_c       \\
              A(E/E_c)^{-5},  &  E_c < E < E_{\rm max}. \\
           \end{array}
        \right.  \label{cascade}
\eeq
The normalization factor is
\beq
          A = \frac{3E_0E_c^{-2}}{7-(E_c/E_{\rm max})^3}.
\eeq

Photon-photon pair production and scattering, and
inverse Compton scattering, are very rapid processes compared to
interactions on matter, due to the large number of photons. When
the photon energies fall below $E_c$
the mean interaction time rises drastically. The thermalization
continues through Compton scattering and pair production in the
field of a nucleus, in a time scale long compared with that of the
cascade.
The pair production cross section is \cite{Lang}
\beq
  \sigma_{\rm pair} = \frac{3\alpha}{8\pi}Z^2\sigma_T
     (\frac{28}{9}\ln\frac{2E}{\me}-\frac{218}{27})
\eeq
and the Compton cross section is ($E\gg\me$)
\beq
   \sigma_C=\frac{3}{8}\sigma_T\frac{\me}{E}(\half+\ln\frac{2E}{\me}).
\eeq

Photons with $E > 19.9\MeV$ disintegrate $\UHe$,
producing $\EHe$, and also D for $E > 26.2\MeV$.
Above the energy $E_{\rm max}$ the cascade
proceeds so rapidly that photodisintegration of nuclei is rare and
can be ignored. The photodisintegration of $\UHe$ begins at $T=$
0.6 keV, when $E_{\rm max}$
becomes larger than the binding energy of $\UHe$. For $T =
0.45$--$0.60\keV$ $\UHe$ photodisintegration produces $\EHe$ (or
$\EH$) only, below $T = 0.45\keV$ also $\D$ is produced, although
with a smaller cross section. The photodisintegration of D begins
earlier, at $T=5.3$ keV, because of the smaller deuteron binding
energy. The $\EHe$ photodisintegration begins at $T = 2.2\keV$,
$\EH$ at $T = 1.9\keV$, and $\ZLi$ at $T = 4.7\keV$.

During the second stage of annihilation, the mean free path of a
photon at a given temperature is always larger than the distance
scale of antimatter regions which annihilate at that temperature.
We can therefore assume that the photons are uniformly distributed
over space.

\subsection{Spectrum of annihilation photons and electrons}

As the temperature
falls the cascade spectrum moves to higher energies and, for
$T\lesssim100\eV$,
it begins to
overlap the initial photon spectrum from annihilation.
Then the lower
part of this initial spectrum is no more converted to a cascade
spectrum before photodisintegration, and
the shape of the initial photon spectrum
becomes important.

In the pion's rest frame its direct decay products, photons, muons, 
and muon neutrinos,
have a single-valued energy, determined by conservation of energy and momentum.
The muon decays via $\mu^{-}\rightarrow e^-+\nu_\mu+\bar\nu_{\rm e}$.
The spectrum of the electron in the muon's rest frame is\cite{Halzen}
\beq
   \frac{dn_e}{dE} = 16\frac{E^2}{m_\mu^3} \left(3-\frac{4E}{m_\mu}\right),
    \quad 0<E<\half m_\mu \quad\hbox{(c.m.)}
\eeq
in the approximation $\me/m_\mu\approx0$.

For the decay products of a moving pion, integration over directions
yields an energy spectrum. The decay
($\pi^\pm\rightarrow\mu^\pm+\nu_\mu$) of a charged pion with velocity
$v_\pi$ and total energy
$E_\pi$  produces a muon with a uniform spectrum in the range
\beq
     \half
     E_\pi\biggl\{\biggl[1+\biggl(\frac{m_\mu}{m_\pi}\biggr)^2\biggr]
     \pm v_\pi\biggl[(1-\biggl(\frac{m_\mu}{m_\pi}\biggr)^2\biggr]\biggr\}.
\eeq
Similarly, the energy of a photon from neutral pion decay
($\pi^0\rightarrow\gamma\gamma$) has a uniform distribution in the range
$\half E_\pi(1\pm v_\pi)$.
For a muon moving with velocity $v$ and energy $E_\mu$, the
electron spectrum becomes
\bea
   \lefteqn{\frac{dn_e}{dE} = } \\
   & &  \frac1{v E_\mu} \left(
               \frac53 -12\frac{E^2E_\mu^2}{m_\mu^4}(1-v)^2
                       +\frac{32}{3}\frac{E^3E_\mu^3}{m_\mu^6}(1-v)^3
                 \right) \nonumber
\eea
for $\half E_\mu(1-v)<E<\half E_\mu(1+v) $, and
\beq
   \frac{dn_e}{dE} =
   \frac{16}{E_\mu} \left(
                 3\frac{E^2E_\mu^2}{m_\mu^4}
                 -\frac43\frac{E^3E_\mu^3}{m_\mu^6}(3+v^2)
                \right)
\eeq
for $0<E<\half E_\mu(1-v)$.

The electrons transfer their energy to background photons
through inverse Compton scattering.
We calculate the scattering rate $R$ for an electron with energy
$E_{\rm e}$ passing through a thermal photon background, in the approximation
$E_{\rm e}\gg m_{\rm e}\gg T$.
Let $E_\gamma$ be the energy transferred from the electron to a photon in one
scattering. Using the Klein-Nishina cross section we get
for a monochromatic photon background
\bea
   \frac{dR}{dE_\gamma} = 6 \frac{n_\gamma}{E_e}\frac{\sigma_{\rm T}}{w} &\Big( &
        \frac14 \big(\frac{1}{1+\epsilon}+1+\epsilon \big)
                      (1-\frac{\epsilon}{w}) +                \label{mono} \\
         && \frac{\epsilon}{w} \ln\frac{\epsilon}{w}
        -(\frac{\epsilon}{w})^2 +\frac{\epsilon}{w}
         \Big)  \theta( 0 < \frac{\epsilon}{w} < 1),  \nonumber
\eea
where $\epsilon \equiv E_\gamma/(E_{\rm e}-E_\gamma)$,
$w \equiv 4E_{\rm e} \epsilon_\gamma/m_{\rm e}^2$, and $\epsilon_\gamma$
is the energy of the background photons.
Integration over the thermal photon spectrum gives the spectrum of
up-scattered photons for one scattering,
\bea
    \frac{dn}{dE_\gamma} &\propto&  \frac{\epsilon^2}{\alpha^3}
        \int_1^\infty \left( e^{(\epsilon/4\alpha)t}-1\right)^{-1} \times
                                                           \label{onescat}       \\
         && \left( \frac14 (\frac{1}{1+\epsilon}+1+\epsilon)(t-1)
                  -\ln t-\frac1t +1 \right) dt,  \nonumber
\eea
where
\beq
     \alpha \equiv \frac{E_{\rm e}T}{m_{\rm e}^2}.
\eeq

The average fractional energy loss
$\langle E_\gamma/E_{\rm e}\rangle$ in one scattering increases with
increasing $\alpha$. At $\alpha\ll1$ the average energy transfer is
$\langle E_\gamma\rangle= 3.60\alpha E_{\rm e}$. At large
$\alpha$ the electron loses most of its energy in one scattering.
At the limit $\alpha\ll1$ the Klein-Nishina cross section reduces into 
the Thomson cross section.

The electron scatters several times, losing a decreasing fraction of 
its energy in each collision. 
The process generates a photon spectrum with most of the photons at low
energies where $dn/dE_\gamma\propto E_\gamma^{-3/2}$.

\begin{figure}[tbh]
\epsfysize=6.6cm
\epsffile{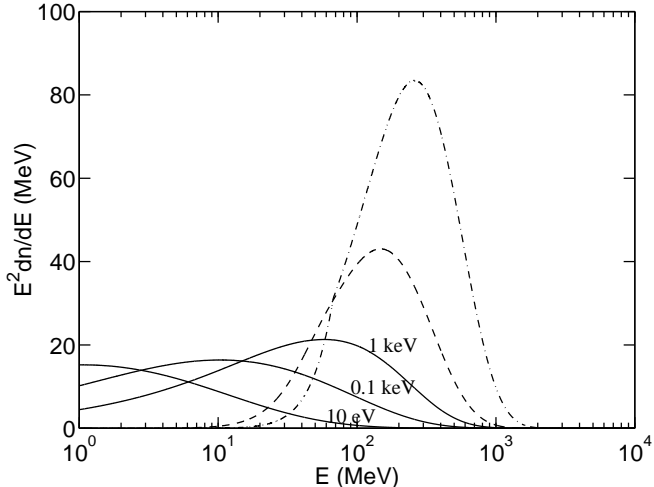}
\caption[a]{\protect
Initial spectra of photons and electrons from pion decay,
for an exponential pion spectrum with the mean energy 329 MeV. 
The {\em dot-dashed line} shows the photon spectrum from the decay 
of a neutral pion.
The {\em dashed line} shows the electron spectrum from the decay 
of a charged pion.
The electrons transfer their energy to background photons through inverse
Compton scattering.
The resulting photon spectra at temperatures 1 keV, 100 eV and 10 eV
are shown by {\em solid lines}.
}
\label{fig:spectra}
\end{figure}

In Fig. \ref{fig:spectra} we plot the spectra of electrons and photons from pion decay,
for an exponential pion spectrum.
We also show photon spectra resulting from inverse Compton scattering.

%

\subsection{Spallation of $\UHe$ by energetic neutrons}

The average energy of a nucleon produced in $\pHe$
annihilation is $\approx 70$ MeV. This is sufficient to
disintegrate a $\UHe$ nucleus. Protons and ions slow down rapidly
compared to the interaction time of nuclear reactions. Neutrons
thermalize much more slowly and may cause significant spallation
of $\UHe$.

Destruction of even a small fraction of $\UHe$ may produce $\EHe$
or D in amounts comparable to the total abundance of these
elements, but destruction of other elements is significant only if
a large fraction of the nuclei is destroyed.  Thus only $\n\UHe$
spallation is important.

For $T < 100$ eV the neutron mean time before spallation becomes
larger than the neutron lifetime and spallation gradually ceases.

\subsection{Lithium}

We do not expect any drastic effects on the $\ZLi$ yield from
antimatter regions. For small scales and large antimatter
fractions the reduction in the $\UHe$ and $\EHe$ yields cause an
even steeper reduction in the $\ZLi$ yield, but the $\UHe$ yield
is a more sensitive constraint.

For large scales, annihilation and photodisintegration of $\ZLi$
is a small effect, just as for $\D$ and $\EHe$, compared to the large
$\EHe$ production from $\UHe$ annihilation and photodisintegration.

Since the standard BBN (SBBN) $\GLi$ yield is much below the $\ZLi$ yield, $\GLi$
production from $\ZLi$ annihilation, spallation, and
photodisintegration could cause a large relative increase in the
$\GLi$ yield.

The $\EH$ and $\EHe$ from photodisintegration and annihilation
have large energies.  They may react with $\UHe$ to produce $\GLi$
and $\ZLi$ before thermalizing.  This nonthermal nucleosynthesis
may proceed via $\EH(\EHe) + \UHe \rightarrow \GLi + \n(\p)$, 
which has a
threshold of 4.80 MeV (4.03 MeV) and is therefore not available
for thermal nucleosynthesis, and it may result in a $\GLi$ yield much
larger than in SBBN\cite{Jed00}.


\section{Numerical Implementation}

\subsection{General}

We use a spherically symmetric geometry where a spherical
antimatter region is surrounded by a thick shell of matter. We assume
equal initial densities $n_b = \nbbar$ in both regions, such that
the average net baryon density $\nbref$ corresponds to
$\langle \eta \rangle = 6\times10^{-10}$.

We give our results as a function of two parameters, the radius of the
antimatter region $r_A$, and the antimatter-matter ratio $R$.
These parameters together with
the net baryon density determine the initial local baryon density
$n_b$ and the volume fraction
$f_V$ covered by antimatter. The volume fraction depends only on 
the antimatter-matter ratio
\beq
     R = \frac{f_V\nbbar}{(1-f_V)n_b} = \frac{f_V}{1-f_V}.
\eeq
The initial baryon density
$n_b$ is linked to the volume fraction through
\beq
   \nbref = (1-f_V)n_b - f_Vn_b.
\eeq
The radius of our grid is $L = r_A/f_V^{1/3}$.
We assume
reflective boundary conditions at the outer boundary of the matter
shell.  This models the situation where antimatter regions of
radius $r_A$ are separated from each other by the distance $2L$
between their centers.

For $R \ll 1$, also $f_V \ll 1$ and $r_A \ll L$, so that we have a
relatively small antimatter region surrounded by a much larger volume
of matter.

The annihilation creates a narrow depletion zone around the
boundary between the matter and antimatter regions.
 An accurate treatment requires a dense
grid spacing in this region. The position of the boundary
moves with time. Therefore a fixed non-uniform grid is not
adequate. We use a steeply non-uniform grid, which is updated at
every time step. The number of grid cells per unit distance is
proportional to the gradient in baryon density. The total number
of cells is kept constant.

We include nucleosynthesis both in matter and antimatter. In
matter we follow the reactions up to $A = 7$, in antimatter up to
$\bar{A} = 4$. Our code includes 15 isotopes: n, p, D, $\EH$,
$\EHe$, $\UHe$, $\GLi$, $\ZLi$, $\ZBe$, $\nbar$, $\pbar$, $\dbar$,
$\tbar$, $\EHebar$, and $\UHebar$. Heavier matter isotopes are
included as sinks.

\subsection{Annihilation and diffusion}

Because of the large uncertainty or lack of data for most of the
relevant annihilation reactions we simply use
\beq
   \langle\sigma v\rangle = \sigma_0
\eeq
for the $\nbar \n$, $\nbar \p$,  $\n\pbar$, 
and all $\nbar \A$, $\n\Abar$
annihilation cross sections and
\beq
   \langle\sigma v\rangle = C(v)\sigma_0
\eeq
for $\pbar \p$ and all $\pbar \A$, $\p\Abar$, and $\Abar \A$
annihilations.
Here
\beq
   C(v) =
   \frac{2\pi|Z_1Z_2|\alpha/v}{1-\exp(-2\pi|Z_1Z_2|\alpha/v)},
\eeq
and we use $\sigma_0 = 40$ mb.  The velocity $v$ in the Coulomb
factor $C(v)$ is the relative velocity, for which we use the
thermal velocity $v=\sqrt{3T/\mu}$, where $\mu$ is the reduced
mass of the annihilating pair.  We also studied the effect of
including an $A^{2/3}$ dependence in $\sigma_0$.

We assume that $\nbar \A$ and $\pbar \A$ have the same nuclear
yields, and that $\n\Abar$ and $\p\Abar$ have the corresponding
antiyields.
The most important $\pbar \A$ reaction is $\pHe$.  For its yield
we use
\bea
   \pbar+\UHe\rightarrow
   0.490{\rm n} & + & 0.309{\rm p} +0.130{\rm D} \nonumber\\
   & + & 0.437\EH +0.210\EHe,
\eea
where we have taken the $\D$, $\EH$, $\EHe$ yields
from\cite{Balestra88},
$\sigma(\pbar\n)/\sigma(\pbar\p) = 0.42$
from\cite{Balestra87}, and we assumed charge exchange has no net
effect,
to get the $\n$ and $\p$ yields.
The $\nbar \A$, $\pbar \A$, $\n\Abar$, and $\p\Abar$ yields for other
nuclei than $\A = \UHe$ are not important.  We estimated yields for
them by assuming that $\pbar(\nbar)$ annihilation is twice as
likely with $\p$ than with $\n$ in the
nucleus\cite{Egidy,Balestra87}, using the experimental $\p$, $\D$,
and $\EH$ yields for $\pbar\GLi$ and $\pbar\ZLi$\cite{Sudov93}, and
otherwise trying to mimic the $\pHe$ data.

There is no data on annihilation of an antinucleus on a nucleus.
For simplicity we assume that the lighter nucleus is annihilated
completely, and the remnants of the heavier nucleus go into $\UHe$
nuclei
and nucleons, with equal number of protons and neutrons.
Especially, annihilation of a nucleus on an antinucleus with equal
mass number leads to total annihilation.

Annihilation, nuclear reactions, and diffusion are
solved together for better accuracy.
Hydrodynamic expansion, spreading of the annihilation yields, and
photodisintegration are treated as separate steps.
We include diffusion of all ions and neutrons.

Annihilation reactions are represented by the differential equation
\beq
   \frac{dY_k}{dt} =
    -\sum_l\nbref \langle\sigma_{kl}^{\rm ann}v\rangle Y_kY_l =
    -\sum_lG_{kl}Y_kY_l, \label{annDE}
\eeq
where the indices $k$ and $l$ refer to
the annihilating isotopes and $Y_k=n_k/\nbref$ 
is the relative abundance.

We integrate Eq. (\ref{annDE}) over the time step $\Delta t$.
We take the implicit equation
\beq
     Y_k^i-Y_k^{i0} = -\sum_l G_{kl}^iY_k^iY_l^i \Delta t, \label{annimp}
\eeq
and linearize it into
\beq
     Y_k^i-Y_k^{i0} = -\sum_l \mu^i G_{kl}^i
     (\hat Y_k^iY_l^i + Y_k^i\hat Y_l^i
     - \hat Y_k^i\hat Y_l^i) \Delta t.
\eeq
Here $Y_k^{i0}$ is the initial abundance and $\hat Y$ is the solution from the previous
iteration step.

We solve this equation iteratively by a modified Newton-Raphson
method.
The ordinary Newton-Raphson method (e.g., \cite{Recipes})
does not work in this case, since it often converges to an unphysical solution
with negative $Y$.  We stabilize the algorithm by introducing
a parameter $\mu$ which initially is set to zero.
We gradually increase the value of $\mu$ between iteration steps,
until the solution has converged and $\mu=1$.

In our code the hydrodynamic expansion is started at a constant
temperature $T=30$ keV. The results are insensitive to the starting
temperature, because late times dominate the expansion.
Equation (\ref{hydroDE}) is solved for $\nbtot$ as an ordinary diffusion
equation. The grid cells are then expanded so that the baryon
distribution corresponds to the solution.

Hydrodynamic expansion is not combined into the same matrix
equation with diffusion and nuclear reactions, but is treated as a
separate step. For this reason the convergence of the code requires a
very small time step at late times, when the hydrodynamic
expansion becomes important.  This limited our ability to
calculate with very large scales.  For our largest scale $r_A =
10^{11}$ m, we did not get a converged result for the CMB
distortion, since it is sensitive to the lowest annihilation
temperatures, although our results for the nuclear
abundances did converge.

\subsection{Spreading of the annihilation products and their reactions}

We model the energy spectrum of a nucleus created in annihilation
by an exponential distribution $\exp(-E/E_0)/E_0$. The spectrum
is cut off at $E=10E_0$. The mean kinetic energy $E_0$ corresponds to
momentum $P_0=350$ MeV$/c$ for neutrons and protons, 
and to $P_0=200$ MeV$/c$ for nuclei with $A>1$.

Consider the spreading of nuclei produced during
one time step, along a linear path. The spherical symmetry allows us to identify paths
with same tangential distance $r_0$ from the symmetry center.
Let $F(E,s,r_0)dr_0$ denote the cumulative spectrum of nuclei at distance $s$ from the tangent point.
The energy
spectrum obeys the differential equation
\bea
   F(E,s,r_0) dr_0 &=&
   F(E-\frac{dE}{ds}ds,s-ds,r_0) dr_0  \label{spr1} \\
   && + F_0(E)g(r) 4\pi r^2dr \frac{d\Omega}{4\pi} . \nonumber
\eea
Here  $g(r)$ is the number of particles
created per unit volume at distance $r$ from the center,
$F_0(E)$ is their initial spectrum, and solid
angle  $\Omega(r_0)=2\pi\sqrt{1-(r_0/r)^2}$ picks directions which 
correspond to the tangential distance $r_0$.

We integrate Eq.~(\ref{spr1}) assuming a reflective boundary condition 
at the outer edge of the grid. Nuclei which fall below 
$E_{\rm min}\sim T$ are considered thermalized. 
The formulae for the energy loss $dE/ds$ due to various scattering 
processes were given in Sec.~II.

Ions lose energy through Coulomb scattering on electrons and ions,
and Thomson scattering
on photons. Neutrons lose energy through scattering on electrons and ions,
or they decay
and thermalize as protons. We include all these effects.
Neutrons are allowed to scatter on an ion once, after which they are stopped.
The strong and weak interaction neutron reactions included are listed in Table I.

\begin{table}[tbh]
\begin{tabular}{ll}
Reaction & Ref. \\ \hline
n+p  total &  \cite{Hale,Lisowski} \\
n+$\pbar$ total & \cite{Mutchler} \\
n+$\UHe$  total & \cite{nalphadata} \\
n+$\UHe\rightarrow\EH+\D$ & \cite{Liskien} (from inverse reaction),
\cite{Tannenwald} \\
n+$\UHe\rightarrow\EH+\p+\n$ & \cite{Tannenwald} \\
n+$\UHe\rightarrow\D+\p+2\n$ & \cite{Tannenwald} \\
n+$\UHe\rightarrow2\D+\n$ & \cite{Tannenwald} \\
n+$\UHe\rightarrow \EHe+2\n$ & \cite{Tannenwald} \\
n $\rightarrow$ p & $\tau$ = 886.7 s \\
\end{tabular}
\caption[a]{\protect
Neutron reactions and references to their cross section data.
}
\end{table}

\subsection{Nonthermal nuclear reactions}

We ignore
spallation of nuclei by energetic nucleons for other nuclei
than $\UHe$. Our results show
that even
$\UHe$ spallation is a relatively small effect, which confirms
that spallation of other nuclei can be safely ignored.

We ignore
in this work also the production of $\GLi$ by non-thermal
$\EHe(\EH)+\UHe$ reactions\cite{Jed00}, but we are
incorporating it for future work\cite{Elina}.

\subsection{CMB distortion}

We calculate the ratio of injected energy to the CMB energy as
\beq
   W = \int^{\rm 2 keV} \frac{1}{\rho_{\rm CMB}(T)}
                        \frac{d\bar\rho_{\rm ann}}{dT} dT
\eeq
and require $W<6\times10^{-5}$ to satisfy the CMB constraint.
Here $\rho_{\rm CMB}$ is the energy density of the background radiation
and $\bar\rho_{\rm ann}$ is the energy density released in annihilation
reactions in form of photons and electrons, averaged over space.
Effectively, we are assuming complete thermalization above
$T=2$ keV (redshift $z\approx 8.5\times 10^6$)
 and no thermalization below it.
We count into $\bar\rho_{\rm ann}$ half of the total annihilation energy.
The other half disappears as neutrinos, and has no effect on
nucleosynthesis or CMB.

\subsection{Photodisintegration}

We compute the initial spectra of electrons and photons from pion decay
following Sec.~IIE.  We assume
an exponential kinetic energy distribution for the pions,
with mean total energy equal to $2m_p/5.7$ = 329 MeV. The electrons transfer
their energy to background photons through inverse Compton
scattering. We compute the spectrum of the upscattered photons
using the Klein-Nishina cross section, assuming a thermal background
spectrum and $E_e \gg m_e$. We then redistribute the energy of the initial
photons
(upscattered and from $\pi_0$ decay),
whose energies are above $E_c$ into the standard cascade
spectrum
[Eq.~(\ref{cascade})].

The photons in this resulting spectrum have then an opportunity to
photodisintegrate.  These photons may pair produce on a nucleus,
Compton scatter, or photodisintegrate nuclei. We allow an
unlimited number of Compton scatterings for a single photon, but
we remove the photon after the production of an e${}^\pm$ pair or a
photodisintegration reaction. The created e${}^\pm$ pairs, as well
as the background electrons which gain energy in Compton
scattering, will produce a second generation of non-thermal
photons by inverse Compton scattering. These secondary photons
are, however, much less energetic than the primary ones, and we
ignore them.

The photodisintegration reactions included in our code are listed in Table II.

\begin{figure}[tbh]
\epsfysize=6.5cm
\epsffile{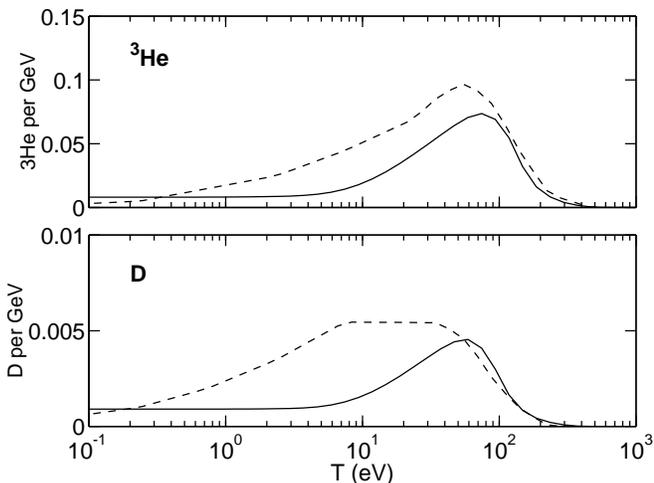}
\caption[a]{\protect
Comparison of our photodisintegration $\EHe$
and D yields (solid lines) with Protheroe et al.\cite{PSB95} (dashed lines).
}
\label{fig:prother}
\end{figure}

In \cite{KS00} we used the results of
Protheroe, Stanev and Berezinsky (PSB)\cite{PSB95} for photodisintegration.
PSB calculated the
amount of $\EHe$ and D produced per 1 GeV of energy released in
the form of photons and electrons, as a function of redshift.
However, their result does not apply for annihilation at low
temperatures, when a significant part of the initial photon spectrum from
annihilation is below
the threshold for photon-photon pair production.
In Fig.~\ref{fig:prother} we compare the PSB yields
with the more detailed treatment
described above which we are now using.

%

We get less $\EHe$ ($\D$)
than the PSB yield  for $T<100\eV$ ($50\eV$).  There are two reasons
for this difference.

\begin{table}[tbh]
\begin{tabular}{ll}
Reaction & Ref. \\ \hline
D+$\gamma \rightarrow$ p+n  & \cite{Hale} (from inverse reaction)
\\
$\EH+\gamma \rightarrow$ D+n & \cite{Faul81,Skopik81} \\
$\EH+\gamma \rightarrow$ p+2n & \cite{Faul81} \\
$\EHe+\gamma \rightarrow$ D+p & \cite{Ticcioni73} \\
$\EHe+\gamma \rightarrow$ 2p+n & \cite{Faul81} \\
$\UHe+\gamma \rightarrow \EHe$+n & \cite{Calarco83} \\
$\UHe+\gamma \rightarrow \EH$+p & same as
$\UHe+\gamma\rightarrow\EHe$+n \\
$\UHe+\gamma \rightarrow $ D+p+n & \cite{Balestra77} \\
\end{tabular}
\caption[a]{\protect
Photodisintegration reactions and references to their cross section data.
}
\end{table}

First, our cross sections (Table II) differ somewhat from what
PSB used.  The main difference is for large photon
energies $E_\gamma \gtrsim 200 \MeV$, where the $\UHe$
photodisintegration cross section
again becomes large and a pion is produced (``pion
photoproduction").
PSB assumed large $\D$ and $\EHe$ yields for these
reactions.  Available data\cite{highphotodata} gives very
small cross sections for the $\UHe+\gamma \rightarrow \EH+\p$,
$\UHe+\gamma \rightarrow \D+\p+\n$, and $\UHe+\gamma \rightarrow
\D+\D$ channels for $E_\gamma > 200\MeV$.  Accordingly, we set the
$\D$ and $\EHe$ yields to zero in this range.  Therefore we get
lower $\D$ and $\EHe$ production at low temperatures as the
cascade moves to these higher energies.

Second, for low temperatures the cascade energies move up to
and beyond the energies of the initial annihilation photons.
Since we only convert to the cascade those initial photons whose
energy is above the cascade turnover $E_c$, our photon spectrum
for photodisintegration does not move up further.  Therefore the
$\D$ and $\EHe$ yields become almost independent of temperature for
$T < 5\eV$.  In our nucleosynthesis runs most of the annihilation
takes place for $T > 5\eV$, but for antimatter regions larger than
$r_A \sim 10^{11}$ m annihilation occurs at these lower
temperatures and
 the photodisintegration contribution should
 become independent of $r_A$.


\section{Results}

\begin{figure}[tbh]
\epsfysize=7.5cm
\epsffile{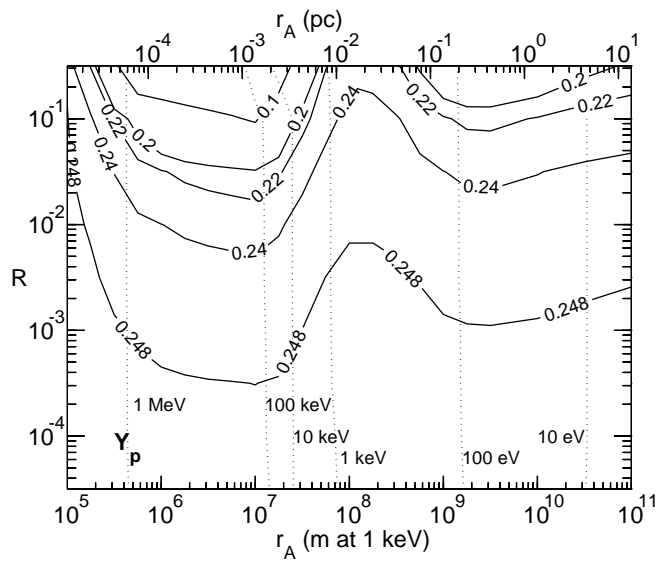}
\caption[a]{\protect
The yield of $\UHe$ as a function of the
antimatter-matter ratio $R$ and the radius $r_A$ of the antimatter regions.
The SBBN result, which is approached in the lower left corner,
is $Y_p = 0.2484$.  The {\em dotted lines} show contours of $T_{\rm
ann}$, the temperature at which half of the antimatter has
annihilated.
}
\label{fig:he4yield}
\end{figure}

\begin{figure}[tbh]
\epsfysize=7.5cm
\epsffile{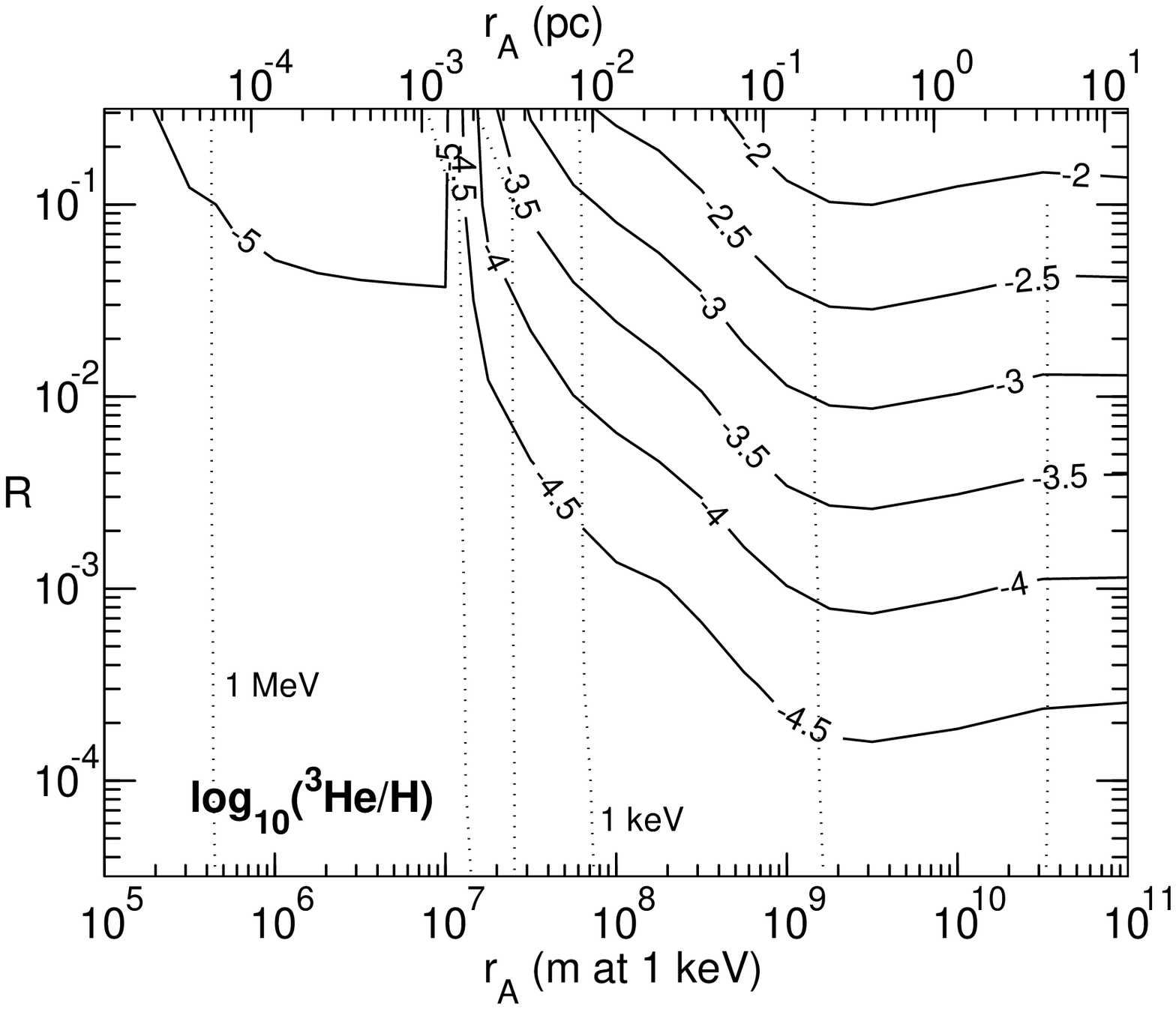}
\caption[a]{\protect
Same as Fig.~\ref{fig:he4yield}, but for $\EHe$.
The SBBN yield is $\EHeH = 1.06\times10^{-5}$.}
\label{fig:he3yield}
\end{figure}

\begin{figure}[tbh]
\epsfysize=7.5cm
\epsffile{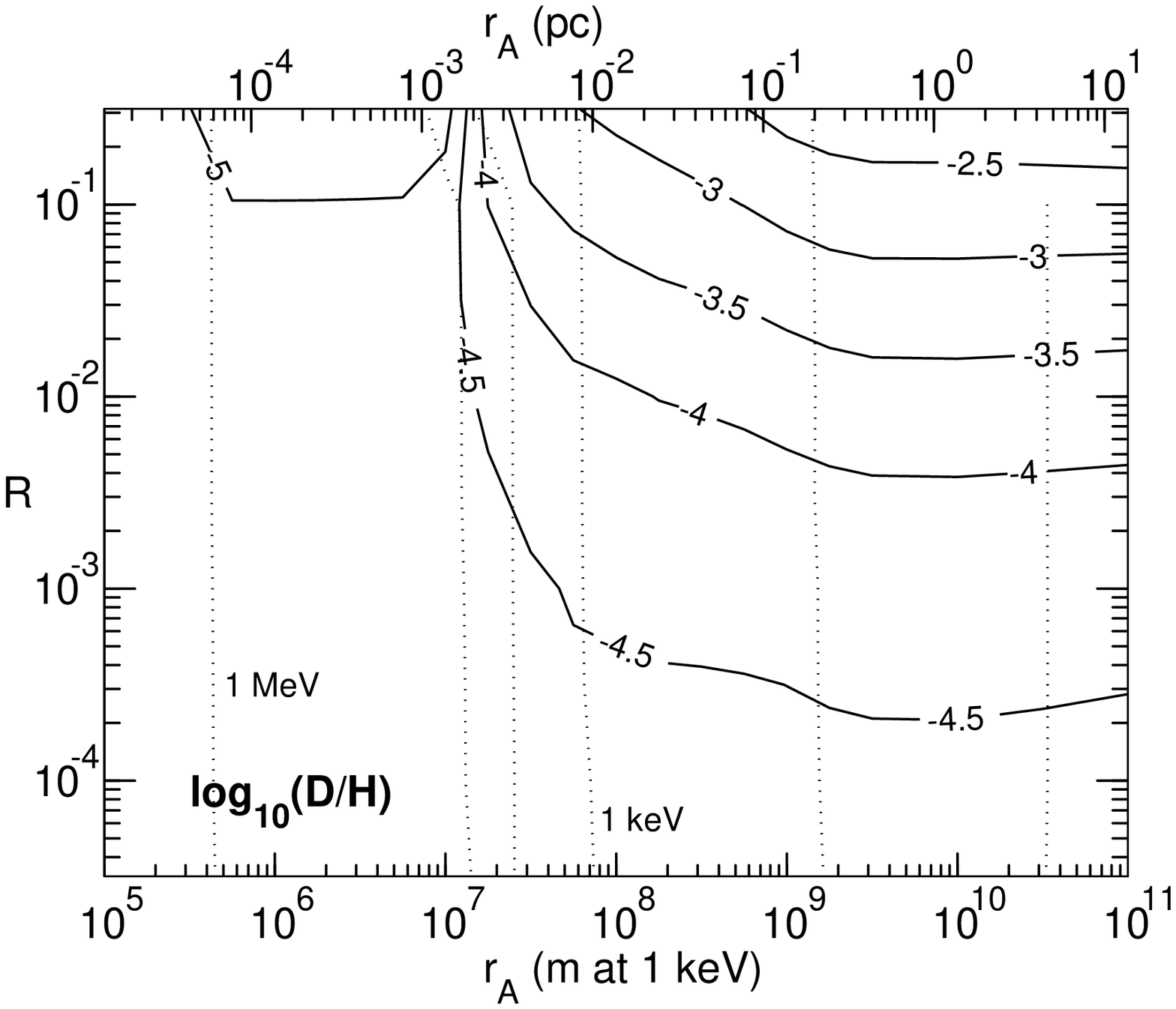}
\caption[a]{\protect
Same as Fig.~\ref{fig:he4yield}, but for $\D$.
The SBBN yield is $\DH = 2.70\times10^{-5}$.}
\label{fig:deuyield}
\end{figure}

 We show light element
yields as a function of the radius $r_A$ of the antimatter region
and the antimatter-matter ratio $R$ in Figs.~\ref{fig:he4yield},
\ref{fig:he3yield}, and \ref{fig:deuyield}.
 We show also the
temperature around which the annihilation is taking place.
All results are for a net baryon density 
$\langle\eta\rangle=6\times10^{-10}$.

For scales smaller than $r_A = 10^5$ m, annihilation happens
before the weak freeze-out, and has no effect on BBN. For scales
between $r_A = 10^5$ m and $r_A = 10^8$ m, neutron annihilation
before $\UHe$ formation leads to a reduction in the $\UHe$ yield compared
to standard BBN (SBBN).

%

%

%

If annihilation is not complete before $\UHe$ formation, it is
delayed significantly because the neutrons have disappeared and
ion diffusion is much slower than neutron diffusion. There will
then be a second stage of annihilation well after nucleosynthesis,
at $T \sim 3$ keV or below.  This
leads to a substantial increase in the yields of $\EHe$ and D.

Antimatter regions in the size range $r_A \sim 10^7$--$10^8$ m are
annihilated in two stages.  In the lower part of this range,
practically all antineutrons diffuse out of the antimatter region
and are annihilated in the first stage, but neutrons diffusing
towards the antimatter region manage to annihilate only an outer
layer of the antiprotons before nucleosynthesis swallows the
remaining neutrons.  Thus the antimatter region that is left for
the second stage of annihilation consists of antiprotons only.

Larger antimatter regions, $r_A \gtrsim 4\times10^7$ m, have also
antineutrons left by the time of nucleosynthesis, and thus
antinucleosynthesis, producing mainly $\UHebar$, takes place
in the antimatter region.  The main
significance of this is that $\pHebar$ annihilation will later produce
high-energy antinucleons, which penetrate deep into the matter
region before annihilating.  Thus not all of the annihilation
occurs in the annihilation zone (``primary" annihilation), but
there is also a significant amount of ``secondary'' annihilation
occurring in a large volume surrounding the annihilation zone.

The main annihilation reaction during the second stage
is $\pHe$.  It produces $\EHe$ and a smaller
amount of D.  Because of their high energy, these annihilation
products penetrate some distance away from the annihilation zone.
Less than half of them end up in the antimatter region and are
annihilated immediately.  The rest end up in the matter region,
but partly so close to the antimatter region that they are
sucked into the annihilation zone and annihilated later (except for the largest
scales studied).

For $r_A \gtrsim 5\times10^7$ m, part of the annihilation
occurs below $T = 0.6\keV$ where $\UHe$ photodisintegration
produces $\EHe$ and D.

\begin{figure}[tbh]
\epsfysize=6.8cm
\epsffile{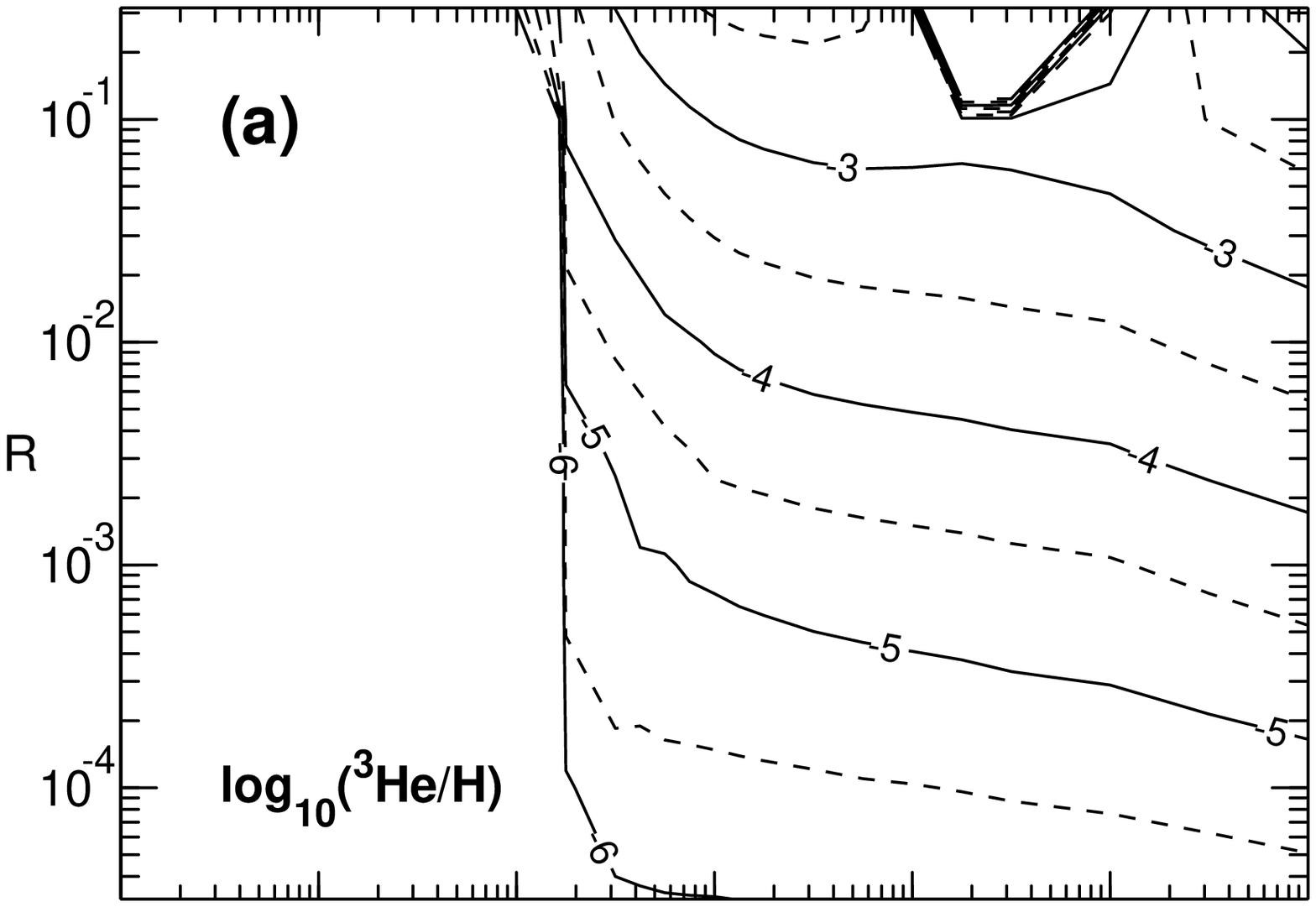}
\epsfysize=6.8cm
\epsffile{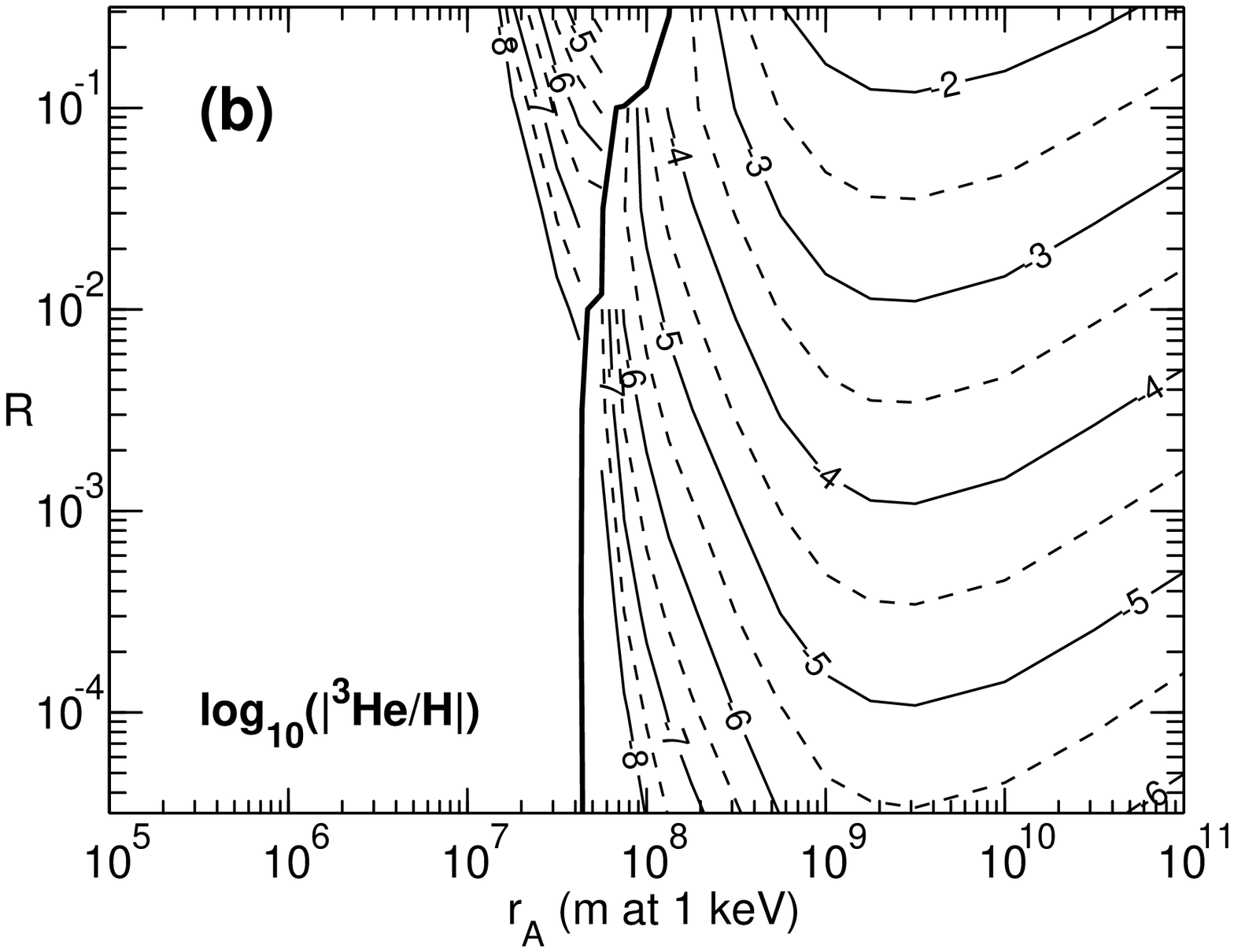}
\caption[a]{\protect
Contribution to the $\EHe$ yield from a) annihilation
and b) photodisintegration. To the left of the {\em thick line} in (b)
the contribution is negative, since $\EHe$ photodisintegration
dominates over photoproduction.
}
\label{fig:he3contr}
\end{figure}

\begin{figure}[tbh]
\epsfysize=6.8cm
\epsffile{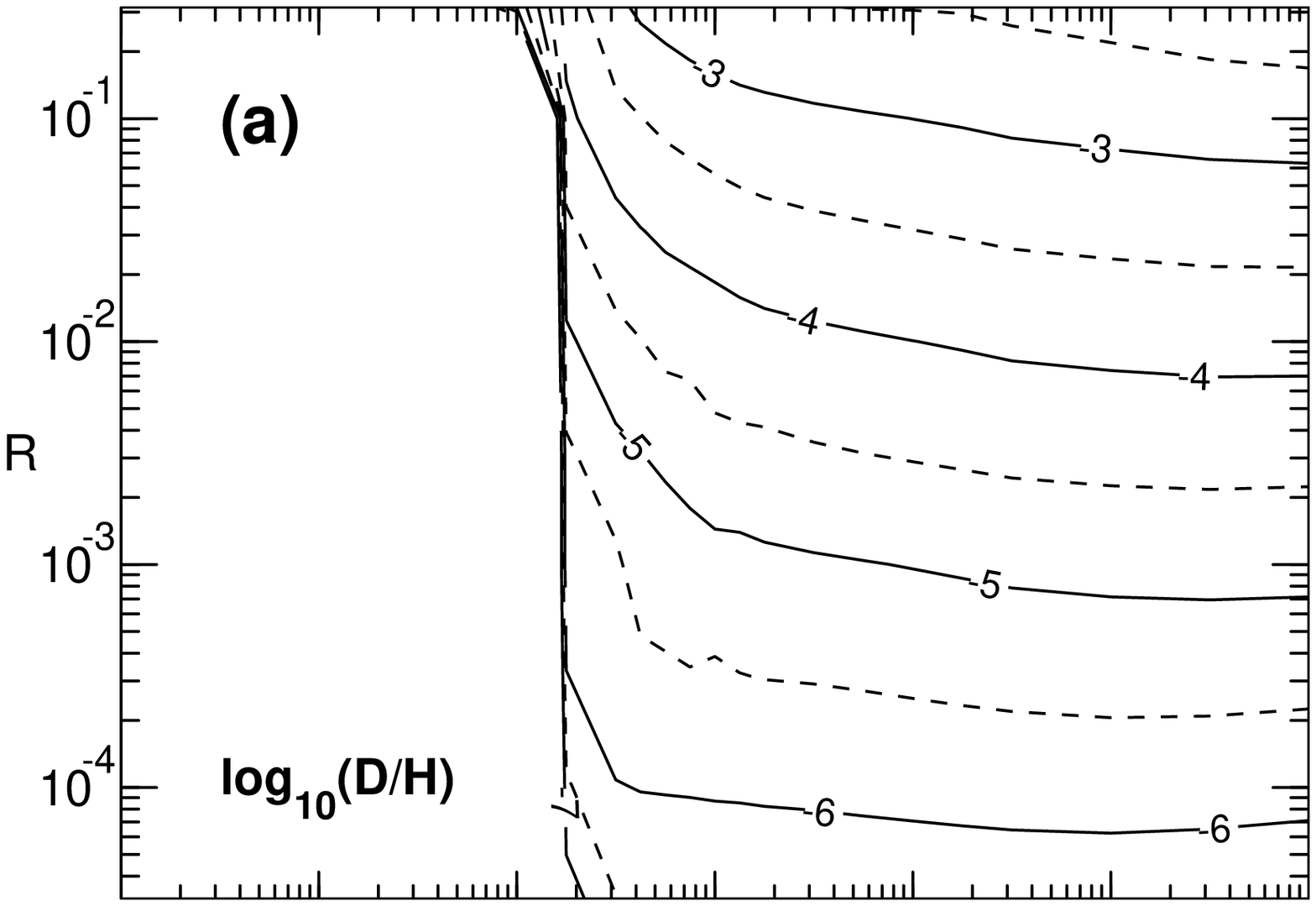}
\epsfysize=6.8cm
\epsffile{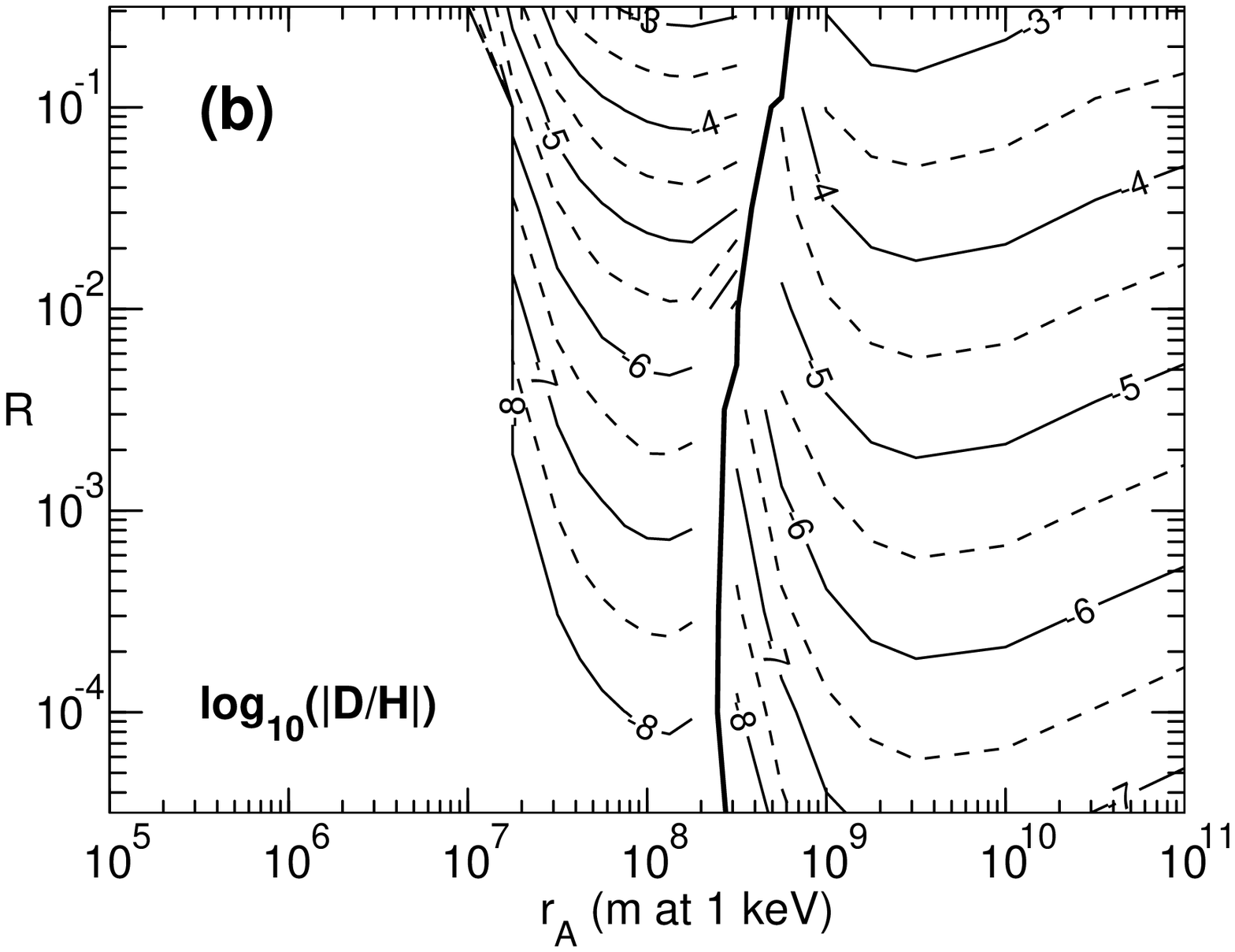}
\caption[a]{\protect
Same as Fig.~\ref{fig:he3contr}, but for the $\D$ yield.
}
\label{fig:deucontr}
\end{figure}

Thus there are two main contributions to $\EHe$ and D production:
annihilation and photodisintegration.  We show these contributions
separately in Figs.~\ref{fig:he3contr} and \ref{fig:deucontr}.

%

Figure \ref{fig:he3contr}a shows the net production of $\EHe$
(including $\EH$) from all annihilation reactions.  The most
important $\EHe$ producing reaction is $\pHe$. Another is $\nHe$,
where the antineutrons come from $\pHebar$ annihilation.  $\EHe$
is destroyed primarily by $\pbar\EHe$ annihilation in the
annihilation zone.

Annihilation production of $\EHe$ increases steeply from $r_A =
2\times10^7$ m to $5\times10^7$ m as a larger part of the
antimatter region survives till the second stage. For $r_A
>5\times10^7$ m, the annihilation production of $\EHe$ keeps
increasing with scale, since the annihilation shifts to lower
temperatures where the $\EHe$ produced in annihilation travels
longer (comoving) distances, and is thus able to better survive
annihilation.

$\EHe$ produced in the matter region by the secondary annihilation
is much more likely to survive and thus this secondary
annihilation produces more, or at least a comparable amount of,
surviving $\EHe$ than the primary
annihilation in the range $r_A \sim 10^8$--$5\times10^9$ m, where most of the $\EHe$
from primary annihilation gets annihilated.
In \cite{KS00} we did not include this secondary annihilation, and
therefore we got a smaller annihilation contribution.

Figure \ref{fig:he3contr}b shows the net production of $\EHe$
from all photodisintegration reactions.
Photodisintegration of $\EHe$ sets in for $r_A \gtrsim 2\times10^7$ m
but has only a small effect.  For $r_A >
5\times10^7$ m, photoproduction of $\EHe$ from $\UHe$ overcomes
$\EHe$ photodestruction and increases up to $r \sim 10^9$ m, as a
larger part of the cascade exceeds the $\UHe$ photodisintegration
threshold.

For $r_A \gtrsim 10^{10}$ m, photoproduction of $\EHe$ decreases
again, as the cascade keeps moving to higher energies.
The photodisintegration
cross sections for $\EHe$ and D production are smaller at these
higher energies, and because the individual photons have higher
energies there are fewer of them.
For even larger scales, $r_A > 10^{11}$ m, we would expect the
photoproduction to stabilize as the cascade gets replaced by the
initial annihilation spectrum (cf.~Fig.~\ref{fig:prother}).

The different dependence of these two contributions on $T$, and
thus on $r_A$, means that annihilation production dominates for
$r_A = 2\times10^7$--$5\times10^8$ m ($T_{\rm ann} > 250\eV$) and
$r_A > 3\times10^{10}$ m ($T_{\rm ann} < 10\eV$), but
photoproduction dominates in the intermediate range
$r_A = 5\times10^8$-- $3\times10^{10}$ m.

In Fig.~\ref{fig:he3contr}a, the feature at $R > 0.1$, $r_A =
10^9$--$10^{10}$ m is due to annihilation of the photoproduced
$\EHe$.

For $\D$ (see Fig.~\ref{fig:deucontr})
we observe the same effects, with some differences.
Annihilation produces about
5 times more $\EHe$ than $\D$, but $\D$ penetrates farther from
the annihilation zone and thus survives better.  Therefore the
$\D$ yield from annihilation is less dependent on $r_A$, as most
of the $\D$ survives already for smaller scales.  The ratio of the
net annihilation production of $\EHe$ and $\D$ is therefore less
than 5, and approaches this number only for the largest scales,
where finally most of the $\EHe$ also survives.

%

Photodisintegration of $\D$ begins already at $T =5.3\keV$,
so it occurs always when the second annihilation stage is reached.
Photoproduction of $\D$ from $\UHe$ can only begin at
$T = 0.45\keV$.  Also the $\D$ yield from $\UHe$ photodisintegration
is less than a tenth of the $\EHe$ yield.  Therefore $\D$
photoproduction overcomes photodisintegration only for scales
$r_A \gtrsim 3\times10^8$ m.

The third significant mechanism for $\D$ and $\EHe$ production
caused by annihilation is spallation of $\UHe$ by the high-energy
neutrons from $\pHe$ annihilation.  For the scales $r_A =
10^7$--$10^8$ m its $\D$ and $\EHe$ yields are about 10\% of
that by annihilation reactions.  For larger scales its relative
importance falls off, as neutrons decay into protons, which are
then thermalized, before encountering a $\UHe$ nucleus.

Because of the large uncertainty about the annihilation cross
sections in reactions involving other nuclei than just nucleons,
we studied the effect of including an $A^{2/3}$ dependence in the
cross section.  This did not have a significant effect on the
primary annihilation in the annihilation zone, but increasing the
$\nHe$ cross section increased the probability of secondary
antineutrons annihilating $\UHe$ instead of protons.  Thus we got
an increased $\EHe$ yield for distance scales $r_A \sim
10^8$--$5\times10^9$ m.  Reducing the $\nHe$ cross section would
have an opposite effect.

\begin{figure}[tbh]
\epsfysize=7.5cm
\epsffile{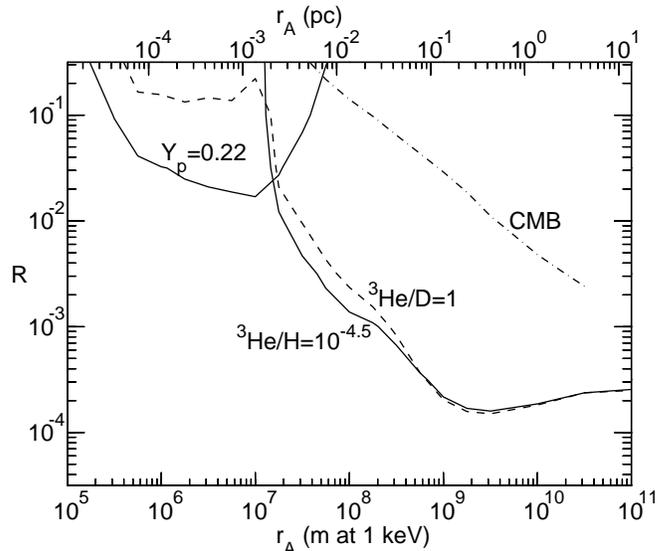}
\caption[a]{\protect
Upper limits from BBN and CMB to the
antimatter-matter ratio $R$ as a function of the radius $r_A$ of the
antimatter regions. The area above the {\em solid lines} is
excluded by $\UHe$ underproduction ($Y_p < 0.22$) or $\EHe$ overproduction
($\EHeH > 10^{-4.5}$).
The {\em dashed line} gives an alternative limit from using $\EHe/\D > 1$
as the criterion for $\EHe$ overproduction.
The {\em dot-dashed line} is the limit from CMB distortion.
}
\label{fig:limits}
\end{figure}

Comparing our calculated yields to
the observed abundances and the primordial abundances derived from
them\cite{Pagel,SBBN}, we obtain upper limits to the amount of antimatter in the
early universe.  We plot the limits from BBN and CMB on the
antimatter-matter ratio $R$ as a function of the radius of the
antimatter region in Fig. \ref{fig:limits}.

For small antimatter regions the limit comes from underproduction
of $\UHe$.
Using $Y_p = 0.22$ as our lower limit to the
primordial $\UHe$ mass fraction, we obtain an upper limit $R
\lesssim 0.02$--0.04 for $r_A = 0.6$--$20\times10^6$ m.  Because
this result is obtained from a calculation with the net baryon
density $\langle\eta\rangle=6\times 10^{-10}$, corresponding to
the SBBN yield $Y_p = 0.248$, a better way to state our $\UHe$
constraint is that we allow a maximum reduction of $\Delta Y_p =
0.028$ from the SBBN result.  Different assumptions on $\eta$ and
observed $Y_p$ could give a smaller acceptable $\Delta Y_p$ and
thus a tighter limit on $R$.  But this does not work in the other
direction, since the $\UHe$ yield falls very rapidly with
increasing $R$.  Thus the limit on $R$ can hardly be relaxed from
our stated value by using different observational constraints.

At larger scales, $r_A>2\times10^7$ m, the limit is set by
overproduction of $\EHe$.  There has been much uncertainty in the
estimated primordial $\EHe$ abundance, because of a large scatter
in its observed abundances and uncertainties about its chemical
evolution\cite{Pagel,Tosi}. Current knowledge suggests a probable primordial
abundance of $\EHeH \sim 10^{-5}$, with three times this
value a reasonable upper limit\cite{Tosi}.
Thus we have used the constraint
$\EHeH<10^{-4.5}$.

The upper limit to $R$ from $\EHe$ falls rapidly as the distance
scale is increased from $2\times10^7$ m to $10^9$ m, where the
limit becomes $R\lesssim2\times10^{-4}$.  For even larger scales
the limit is slightly relaxed but stays below $3\times10^{-4}$.

Fig.~\ref{fig:limits} can be compared to Fig.~2 of Rehm and
Jedamzik\cite{RJ98} or to Fig.~2 of \cite{KS00}. Our $\UHe$ yield
is slightly larger and the corresponding limit to  $R$
 weaker than
in \cite{RJ98}, because our net baryon density, $\langle\eta\rangle=6\times
10^{-10}$, is larger than the one used in \cite{RJ98},
$\langle\eta\rangle=3.43\times 10^{-10}$. Near $r_A \sim 10^8$ m we now
get a tighter limit on $R$ due to a higher $\EHe$ yield than we
gave in \cite{KS00}.  This is due to $\EHe$ production by
secondary annihilation in the matter region, which was ignored in \cite{KS00}.

%

These limits are stronger than those from
the CMB spectrum distortion for scales $r_A \leq 10^{11}$ m. We
did not calculate the yields for larger scales, but the $\EHe$ and
$\D$ yields should become roughly independent of $r_A$, since for
these larger scales the primary annihilation products penetrate
far enough from the annihilation region to survive, and the
spectrum responsible for photodisintegration is the initial
annihilation spectrum, so the dependence on the annihilation
temperature disappears. The CMB limit should then become stronger
than the $\EHe$ constraint near the scale $r_A \sim 10^{12}$ m.


\section{Conclusions}

We have studied the effect of antimatter regions of a comoving
size $r_A \sim 10^{-5}$--$10$ pc on big bang nucleosynthesis.
Smaller antimatter regions annihilate before weak freeze-out and
are not likely to lead to observable consequences.  Larger regions
annihilate close to, or after recombination, and the amount of
antimatter in such regions is tightly constrained by the CMB and
CDG spectra.

Regions smaller than $r_A \sim 2\times10^{-3}$ pc annihilate before
nucleosynthesis.  The annihilation occurs due to neutron and
antineutron diffusion and leads to a reduction in the $n/p$ ratio
and thus to a reduction in $Y_p$.  Requiring $Y_p \geq 0.22$, we
obtain an upper limit $R \lesssim$ few \% for the primordial
antimatter-matter ratio for antimatter regions in the size range
$r_A \sim (0.1$--$2)\times10^{-3}$ pc.

If the annihilation is not complete by nucleosynthesis, at $T \sim
80$ keV, it is significantly delayed, since all neutrons and
antineutrons are incorporated into $\UHe$ and $\UHebar$, and
(anti)proton diffusion is much slower.  There will be a second
stage of annihilation at $T \lesssim 3$ keV, when proton and ion
diffusion finally become effective in mixing the remaining
antimatter with matter.

This second stage of annihilation leads to production of a large
amount of
$\EHe$ and a smaller amount of D, through several mechanisms.

Annihilation of $\UHe$ with antiprotons in the annihilation zone
separating the matter and antimatter region produces $\EHe$ and
$\D$ which are deposited some distance away from the annihilation
zone.  A large fraction of these annihilation products gets
however sucked into the annihilation zone later and is thus
annihilated.  The surviving fraction increases with increasing
distance scale, since this corresponds to a decreasing
annihilation temperature.  At lower temperatures the energetic
ions from annihilation penetrate larger (comoving) distances into
the matter region.

Antinucleosynthesis in the antimatter region produces $\UHebar$,
whose annihilation produces antinucleons and smaller antinuclei.
Of these, especially the antineutrons penetrate deep into the
matter region, where they can annihilate $\UHe$ producing $\EHe$,
which has now a much better chance to survive.

An important source of $\EHe$ and D is photodisintegration of
$\UHe$ by the annihilation radiation.  The large energy part of
the initial radiation spectrum is converted into an
electromagnetic cascade spectrum.  The large-energy cut-off of the
cascade exceeds the $\UHe$ photodisintegration threshold when the
temperature has fallen below 0.6 keV.  For $r_A \gtrsim
3\times10^{-2}$ pc most of the annihilation occurs below this
temperature and thus the photoproduction of $\EHe$ and D becomes
important.
Photodisintegration of $\UHe$ is the dominant source of $\EHe$ for
$r_A \sim $ 0.1--10 pc.  For larger distance scales the
annihilation mainly occurs at lower temperatures where the photon
spectrum shifts to higher energies where it causes less
photodisintegration.

Another source of $\EHe$ and D is the spallation of $\UHe$ by
high-energy neutrons from annihilation reactions. This effect
is at least an order of magnitude smaller than the ones discussed above.

$\EHe$ and D are also destroyed by photodisintegration, but since
the total amount of these isotopes is much less than that of
$\UHe$, this is a small effect.

For scales larger than $r_A \sim 2\times10^{-3}$ pc the tightest
constraint on the primordial amount of antimatter is due to $\EHe$
overproduction. $\pHe$ annihilation produces several times
more $\EHe$ than D and $\UHe$ photodisintegration produces over 10
times more $\EHe$ than D. For scales larger than $r_A \sim 0.1$ pc,
the requirement $\EHeH < 10^{-4.5}$ gives an
upper limit $R \lesssim 3\times10^{-4}$.

Rehm and Jedamzik\cite{Rehm00} have studied this same problem and
obtained results that seem to be in qualitative agreement with ours,
but they find a lower $\EHe$ yield.  Their upper limit to $R$
from $\EHe$ overproduction is weaker than ours by about a factor of 2.
They also criticize our use
of $\EHeH$ as a constraint.  Therefore we show in
Fig.~\ref{fig:limits} also the constraint $\EHe/\D < 1$, which is
observationally more secure\cite{Sigl95}.  As can be seen
from the figure, the limits to $R$ stay essentially the same.
By assuming that the low $\GLi$/H observed in some Population II and
disk stars is an upper limit to its primordial value, they
obtain an even tighter limit on $R$ from $\GLi$
overproduction\cite{Rehm00}. However, $\GLi$ is very
fragile and is thus likely to be
depleted in these stars.  
The main source of $\GLi$ is spallation by cosmic
rays in the interstellar medium. Thus the primordial abundance of
$\GLi$ based on observations is very uncertain, 
as noted also in \cite{Rehm00}, and could be much
lower or much higher than the one observed.

In conclusion, we have established nucleosynthesis constraints
on the amount of antimatter in the early universe which are
tighter, by a large factor, than
those from the CMB spectrum, or any other known observational
constraint, for antimatter regions smaller than $\sim 10$ pc.

\section*{Acknowledgements}

We thank T.~von Egidy, A.M.~Green, K.~Jedamzik, K.~Kajantie,
P.~Ker\"{a}nen, D.P.~Kirilova, J.~Rehm, J.-M.~Richard, M.~Sainio,
M.~Shaposhnikov, G.~Steigman, M.~Tosi, and S.~Wycech for
discussions. We thank the Center for Scientific Computing
(Finland) for computational resources.

\end{document}